\documentclass[12pt,fleqn]{article}
\usepackage{amsmath}
\usepackage{slashed}
\usepackage{url}
\usepackage{bm}
\usepackage{graphicx}
\usepackage[utf8]{inputenc}
\usepackage[english]{babel}
\usepackage{amsmath}
\usepackage{amsfonts}
\usepackage{amssymb}
\usepackage{graphicx}
\usepackage{float}
\usepackage{hyperref}
\begin{document}

\title{The dual behavior of quantum Fields and the big Bang}
\author         {Malik Matwi}
\date{Damascus University, Syria, 2014-2015\\
malik.matwi@hotmail.com}
\maketitle
\tableofcontents

\begin{abstract}
We modify the propagation of the quantum fields like 1.1 and 1.2 for the quarks and gluons. With that we have finite results(without ultra violet divergence) in the perturbation theory. Then we search for $\rm{a}^2p^2\rightarrow0 $ and $ \rm{a}^2k^2\rightarrow0$  with fixing the Lagrange parameters $Z_i$, therefore we can ignore our modification. We find the situation $\rm{a}^2p^2\rightarrow0 $ and $ \rm{a}^2k^2\rightarrow0$ associates with the free particles situation $g\rightarrow0$(g is the coupling constant) and the situation $\rm{a}\neq0$ associates with the perturbation breaking.
We try to give the modification terms  $\rm{a}^2p^2 /(1+ \rm{a}^2p^2) $ and $ \rm{a}^2k^2 /(1+ \rm{a}^2k^2)$  physical aspects, for that we find the corresponding terms in the Lagrange. To do that we find the role of those terms in the Feynman diagrams, in self energies, quarks gluons vertex,\ldots.
We see we can relate the propagation modification to fields dual behavior, pairing particle-antiparticle appears as scalar particles with mass $1/\rm{a}$ (chapter 2). For the quarks we can interrupt  these particles as pions with charges -1, 0, +1 .
If we used the propagation modification for deriving the quarks static potential U(r) of exchanged gluons and pions we find  $U(0)\sim 1/\rm{a}$ if we compare this with the Coulomb potential we find the length a equivalents to smallest distance between the interacted quarks.
We use the static potential in quarks plasma study. We find the free and confinement quarks phases. We suggest a nuclear compression. We find there is a decrease in the global pressure due to the nuclear condensation. We use this decrease in Friedman equations solutions, we find we can control the dark matter and dark energy, we can cancel them.
\end{abstract}
The key words: propagation modification, Lagrange parameters, quarks static potential, fields dual behavior, quarks plasma phase, quarks condensation phase, nuclear compression, the Big Bang, controlling dark energy and dark Matter.

\section{Quarks and Gluons propagation modification}
To remove the UV divergences in the Quarks and Gluons perturbation interaction, we modify the propagation like:
\begin{equation}\numberwithin{equation}{section}
\overline \Delta  _{\mu \nu }^{ab} (k^2 ) = \frac{{g_{\mu \nu } \delta ^{ab} }}{{k^2  - i\varepsilon }}\left( {1 - \frac{{{\rm{a}}^{\rm{2}} k^2 }}{{1{\rm{ + a}}^{\rm{2}} k^2 }}} \right)\text{for glouns }
\end{equation}
\begin{equation}
\bar S_{ij} (\slashed{p}) = \frac{{ - \slashed{p}\delta _{ij} }}{{p^2  - i\varepsilon }}\left( {1 - \frac{{{\rm{a}}^2 p^2 }}{{1{\rm{ + a}}^2 p^2 }}} \right)\text{for quarks}
\end{equation}
the indexes a and b are gluons indexes, i and j color indexes and a is critical length, $\hbar=c=1$.
We use this modification in calculation the quarks self-energy for the perturbation interaction with the gluons, then we renormalize the interaction and
search for the condition $\rm{a}^2p^2\rightarrow0$ and $\rm{a}^2k^2\rightarrow0$ . we have
\begin{figure}[h!]
  \includegraphics[width=0.55\textwidth]{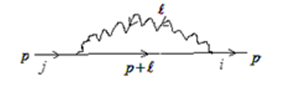}
  \caption{The quarks self energy in strong interaction.}
\end{figure}
\begin{align*}\numberwithin{equation}{section}
i\Sigma _{ij} (\slashed{p})& = \int {\frac{{d^4 \ell }}{{(2\pi )^4 }}[ig_s {\rm{\gamma }}^\mu  T_{ik}^a } \frac{{\overline S _{kl} ( \slashed{p}+\slashed{\ell} )}}{i}ig_s {\rm{\gamma }}^\nu  T_{lj}^b ]\frac{{\overline \Delta  _{_{\mu \nu } }^{ab} (\ell ^2 )}}{i}
\\
&= g_s^2 T_{ik}^a T_{lj}^b \int {\frac{{d^4 \ell }}{{(2\pi )^4 }}[{\rm{\gamma }}^\mu  } \frac{{( -\slashed{p}-\slashed{\ell}) \delta _{kl} }}{{(p + \ell )^2 }}{\rm{\gamma }}^\nu  ]\frac{{g_{\mu \nu } \delta ^{ab} }}{{\ell ^2 }}
\end{align*}
So
\begin{align*}
i\Sigma _{ij} (\slashed{p}) &= g_s^2 T_{ik}^a T_{kj}^a \int {\frac{{d^4 \ell }}{{(2\pi )^4 }}[{\rm{\gamma }}^\mu  } \frac{{( -\slashed{p}-\slashed{\ell})}} {{(p + \ell )^2 }}{\rm{\gamma }}^\nu  ]\frac{{g_{\mu \nu } }}{{\ell ^2 }} \\
&= g_s^2 C(R)\delta _{ij} \int {\frac{{d^4 \ell }}{{(2\pi )^4 }}[{\rm{\gamma }}^\mu  } \frac{{(-\slashed{p}-\slashed{\ell})}}{{(p + \ell )^2 }}{\rm{\gamma }}_\mu  ]\frac{1}{{\ell ^2 }}{\rm{}}
\end{align*}
using ${\rm{ \gamma }}^\mu  (-\slashed{p}-\slashed{\ell} ){\rm{\gamma }}_\mu= 2( -\slashed{p}-\slashed{\ell} )$, it becomes
\begin{equation*}
i\Sigma _{ij} (\slashed{p}) = 2g_s^2 C(R)\delta _{ij} \int {\frac{{d^4 \ell }}{{(2\pi )^4 }}\frac{{(-\slashed{p}-\slashed{\ell})}}{{(p + \ell )^2 }}} \frac{1}{{\ell ^2 }}{\rm{ }}
\end{equation*}

Now we use the gluon modified propagation
\begin{equation*}
\overline \Delta  _{\mu \nu }^{ab} (k^2 ) = \frac{{g_{\mu \nu } \delta ^{ab} }}{{k^2  - i\varepsilon }}\left( {1 - \frac{{{\rm{a}}^{\rm{2}} k^2 }}{{1{\rm{ + a}}^{\rm{2}} k^2 }}} \right){\rm{}}
\end{equation*}
we get
\begin{align}
i\Sigma _{ij} (\slashed{p}) &= 2g_s^2 C(R)\delta _{ij} \int {\frac{{d^4 \ell }}{{(2\pi )^4 }}} \frac{{(-\slashed{p}-\slashed{\ell})}}{{(p + \ell )^2 }}\frac{1}{{\ell ^2 }}\left( {1 - \frac{{{\rm{a}}^{\rm{2}} \ell ^2 }}{{1{\rm{ + a}}^{\rm{2}} \ell ^2 }}} \right)
\\
&= 2g_s^2 C(R)\delta _{ij} \int {\frac{{d^4 \ell }}{{(2\pi )^4 }}} \frac{{(-\slashed{p}-\slashed{\ell})}}{{(p + \ell )^2 }}\frac{1}{{\ell ^2 }}\frac{1}{{1{\rm{ + a}}^{\rm{2}} \ell ^2 }}
\end{align}

For massive quarks, the self-energy becomes:
\begin{equation*}
i\Sigma _{ij} (\slashed{p}) = g_s^2 C(R)\delta _{ij} \int {\frac{{d^4 \ell }}{{(2\pi )^4 }}} \frac{N}{{(p + \ell )^2  + m_q^2 }}\frac{1}{{\ell ^2+ m_\gamma ^2 }}\frac{1}{{1{\rm{ + a}}^{\rm{2}} \ell ^2 }}
\end{equation*}
with ${\rm{ }}N = {\rm{\gamma }}^\mu  ( -\slashed{p}-\slashed{\ell}+ m){\rm{\gamma }}_\mu$, using the Feynman formula:
\begin{align*}
&\frac{1}{{((p + \ell )^2 {\rm{ +  }}m^2 ) \cdot (\ell ^2 {\rm{ +  }}m_\gamma ^2 ) \cdot (1/{\rm{a}}^{\rm{2}}  + \ell ^2 )}} \\
&= \int {dF_3 \frac{1}{{\left[ {((p + \ell )^2 {\rm{ +  }}m^2 )x_1  + (\ell ^2 {\rm{ +  }}m_\gamma ^2 )x_2  + (1/{\rm{a}}^{\rm{2}}  + \ell ^2 )x_3 } \right]^3 }}}
\end{align*}

with $\int {dF_3 } = 2\int\limits_0^1 {dx_1 dx_2 dx_3 \delta (x_1+x_2 {\rm{ + }}x_3  - 1)}$\\
and setting the transformation $q=\ell+x_1p$
with changing the integral to be over q and making transformation to Euclidean space, the self-energy becomes[2]
\[
i\Sigma _{ij} (\slashed{p}) = g_s^2 C(R)\delta _{ij} i\int {\frac{{d^4 \bar q}}{{(2\pi )^4 }}} \frac{1}{{{\rm{a}}^{\rm{2}} }}\int {dF_3 \frac{N}{{[\bar q^2  + D]^3 }}}
\]
with $D=-x_1^2 p^2+x_1 p^2+x_1 m^2+x_2+{{\rm{\emph{m}}}_\gamma ^2 }+(1-x_1-x_2)1/\rm{a}^2$\\
The linear term in q integrates to zero, using $q=\ell+x_1p$, N is replaced with[2]
\[
N \to -2(1-x_1 )\slashed{p}-4m
\]
Using the relation
\[
\int {\frac{{d^d \bar q}}{{(2\pi )^d }}\frac{{(\bar q^2 )^a }}{{(\bar q^2  + D)^{\rm{b}} }}}  = \frac{{\Gamma ({\rm{b}} - a - \frac{d}{2})\Gamma (a{\rm{ + }}\frac{d}{2})}}{{(4\pi )^{\frac{d}{2}} \Gamma ({\rm{b}})\Gamma (\frac{d}{2})}}D^{ - {\rm{(b}} - a - \frac{d}{2})}\]
The integral over q in Euclidean space becomes:
\begin{align*}
\Sigma _{ij} (\slashed{p}) &= g_s^2 C(R)\delta _{ij} \frac{1}{{{\rm{a}}^{\rm{2}} }}\int {dF_3 N\frac{{\Gamma (3{\rm{ - }}2)\Gamma (2)}}{{(4\pi )^2 \Gamma (3)\Gamma (2)}}} D^{ - (3 - 2)}\\
&= g_s^2 C(R)\delta _{ij} \frac{1}{{{\rm{a}}^{\rm{2}} }}\int {dF_3 \frac{N}{{16\pi ^2  \cdot 2}}} D^{ - 1}
\end{align*}
The self-energy becomes
\[
\Sigma _{ij} (\slashed{p}) = g_s^2 C(R)\delta _{ij} \frac{1}{{{\rm{a}}^{\rm{2}} }}\int\limits_0^1 {dx_1 \int\limits_0^{1 - x_1 } {dx_2 \frac{N}{{16\pi ^2 }}} } \frac{1}{D}\]
\[
= \frac{ g_s^2 C(R)\delta _{ij}}{{16\pi ^2 }}\int\limits_0^1 {dx_1 \int\limits_0^{1 - x_1 } {dx_2 } } \frac{{ - 2(1 - x_1 )\slashed{p} - 4m}}{{{\rm{a}}^{\rm{2}} \left[ { - x_1^2 p^2  + x_1 p^2  + x_1 m^2  + x_2 m_\gamma ^2  + (1 - x_1  - x_2 )\frac{1}{{{\rm{a}}^{\rm{2}} }}} \right]}}
\]
we write
\begin{equation}
\Sigma _{ij} (\slashed{p}) = C(R)\delta _{ij} \frac{{g_s^2 }}{{8\pi ^2 }}\int\limits_0^1 {dx_1 \int\limits_0^{1 - x_1 } {dx_2 } } \frac{{ - (1 - x_1 )\slashed{p} - 2m}}{{\left[ {{\rm{a}}^{\rm{2}} f + (1 - x_1  - x_2 )} \right]}}
\end{equation}
with $f =  - x_1^2 p^2  + x_1 p^2  + x_1 m^2  + x_2 m_\gamma ^2 $ \\
this is finite result(without diverges).\\
Now we renormalize the fermions propagation to give the real states and let $\rm{a}\rightarrow0$. \\
The interacted quarks propagation becomes[2]:
\begin{equation}
\overline S (\slashed{p})^{ - 1}  = \slashed{p} + m - \Sigma (\slashed{p})
\end{equation}
To renormalize the interacted field, we write it like
\begin{equation}
\overline S (\slashed{p})^{ - 1}  =  \slashed{p}+ m - \Sigma (\slashed{p}) = Z_2\slashed{p}  + Z_m m
\end{equation}
The parameters $Z_2$ and $Z_m$ are the renormalization parameters, latter we try to make them constants. For the interacted field $\psi$  we have:
\[
\left\langle 0 \right|\psi (\slashed{p})\bar \psi ( -\slashed{p} )\left| 0 \right\rangle  = \frac{1}{i}\frac{1}{{ \slashed{p}+ m - \Sigma (\slashed{p})}} = \frac{1}{i}\frac{1}{{Z_2\slashed{p}  + Z_m m}} = \frac{1}{{iZ_2 }}\frac{1}{{ \slashed{p}+ Z_2^{ - 1} Z_m m}}
\]
We can rewrite
\[
\left\langle 0 \right|\sqrt {Z_2 } \psi (\slashed{p})\sqrt {Z_2 } \bar \psi ( -\slashed{p} ){\rm{ }}\left| 0 \right\rangle  = \frac{1}{i}\frac{1}{{ \slashed{p}+ Z_2^{ - 1} Z_m m}}
\]
And make $m_0  = Z_2^{ - 1} Z_m m$ and $\psi _0 {\rm{ = }}\sqrt {Z_2 } \psi$
with that we have bare fields $\psi_0$  they are like the free fields and like the classical fields, so we can make them independent on the interaction, so
$\partial\psi_{0}/\partial{p^2}=\partial{m_{0}}/\partial{p^2}=0$  for $\rm{a}\rightarrow0$ \\
by that we renormalize the interaction. We make $\psi$  the interacted field with mass m  the physical mass, but we have to make $Re\Sigma(-m)=0$ in (1.6) but with $m_\gamma ^2< 0$ .
From 1.5 and 1.7 we have
\[
Z_2  = 1 + C(R)\frac{{g_s^2 }}{{8\pi ^2 }}\int\limits_0^1 {dx_1 \int\limits_0^{{\rm{   }}1 - x_1 } {dx_2 } } \frac{{1 - x_1 }}{{\left[ {{\rm{a}}^{\rm{2}} f + (1 - x_1  - x_2 )} \right]}}
\]
\[
Z_m  = 1 + C(R)\frac{{g_s^2 }}{{8\pi ^2 }}\int\limits_0^1 {dx_1 \int\limits_0^{{\rm{   }}1 - x_1 } {dx_2 } } \frac{2}{{\left[ {{\rm{a}}^{\rm{2}} f + (1 - x_1  - x_2 )} \right]}}\] and $f =  - x_1^2 p^2  + x_1 p^2  + x_1 m_q^2  + x_2 m_\gamma ^2$

By that we removed the self-energy of the interacted quark and make the mass varies.
For easy we ignore $m_q$ and $m_\gamma$ so
\[
Z_2 = 1 + C(R)\frac{{\alpha _s }}{{2\pi }}\int\limits_0^1 {(1 - x)\ln \left( {1 + \frac{1}{{{\rm{a}}^2 p^2 x}}} \right)dx}\]
\[= 1 + \frac{{C(R)\alpha _s }}{{4\pi ({\rm{a}}^2 p^2 )^2 }}\left[ {({\rm{a}}^2 p^2 )^2 \ln \left( {1 + \frac{1}{{{\rm{a}}^2 p^2 }}} \right) - {\rm{a}}^2 p^2  + (2{\rm{a}}^2 p^2  + 1)\ln ({\rm{a}}^2 p^2  + 1)} \right]
\]
\\
Now we fix $Z_2=constant$ and search for the situations ${\rm{-a}}^2p^2\rightarrow0$ for timelike and ${\rm{a}}^2p^2\rightarrow0$ for spacelike, we have
\[
\frac{{\alpha _s }}{{({\rm{a}}^2 p^2 )^2 }}\left[ {({\rm{a}}^2 p^2 )^2 \ln \left( {1 + \frac{1}{{{\rm{a}}^2 p^2 }}} \right) - {\rm{a}}^2 p^2  + (2{\rm{a}}^2 p^2  + 1)\ln \left( {{\rm{a}}^2 p^2  + 1} \right)} \right] = {\rm{c}}
\]
For spacelike $p^2>0$ \\
\begin{figure}[h!]
  \includegraphics[width=0.45\textwidth]{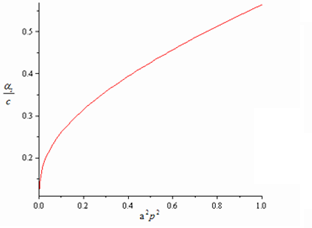}
  \caption{The behavior of the length a with fixing $Z_2$.}
\end{figure}
According to this figure we have
$
{\rm{a}}^2 p^2  = e^{\frac{{ - c}}{{\alpha _s }}}  \to 0{\rm{ }}$ when ${\rm{ }}\alpha _s  \to 0$  this is the decoupling; ${\rm{ }}p^{\rm{2}}  >  > \Lambda _{QCD}^2 {\rm{ }}$.\\
It is the free quarks and glouns situation; $\alpha_s\rightarrow0$ occurs at high energy for free quarks phase.\\
Because $p\rm{a}\rightarrow0$ so $p<<1/\rm{a }$ this gives $r>>\rm{a}\rightarrow0$, therefore the propagation modification is ignored.
So the behavior of the length a is like the behavior of the coupling constant $\alpha_s$ and the modification terms are removed $p\rm{a}<<1$ at high energy(free quarks phase).\\

For the limited low energy we fix ${\alpha_s}/{\rm{a}}^2=constant*\sigma$, $\sigma$ is string tension appears in the low energy static potential $U(r)$ as we will see, for $\rm{a}\rightarrow0$ we have
\[Z_2  = 1 + C(R)\frac{{\alpha _s }}{{4\pi }}\left( {\frac{3}{2} - \ln \left( {p^2 {\rm{a}}^2 } \right) + O(p^2 {\rm{a}}^2 )} \right) \to 1,\text{when }{\rm{ a}} \to {\rm{0}}
\]
\[Z_m  = 1 + C(R)\frac{{\alpha _s }}{\pi }\left( {1 - \ln \left( {p^2 {\rm{a}}^2 } \right) + O(p^2 {\rm{a}}^2 )} \right) \to 1,\text{when }{\rm{ a}} \to {\rm{0}}\]

We know the strong interaction coupling constant ${\alpha_s}$ extremely increases at low limited energy, therefore, according to the figure, we can't let $\rm{a}\rightarrow0$, so we assume when the perturbation breaks down the length a could not be removed and takes non-zero value let it $\rm{ a}_0$, so the propagation modification takes place.\\

\begin{flushleft}
\emph{The confinement situation:}
\end{flushleft}

According to the figure(2) it is possible to have ${\rm{a}}p>1$(the coupling constant $\alpha_s$ extremely increases at low energy), therefore $p>1/{\rm{a}}\rightarrow r<\rm{a}$ which is the quarks confinement phase at low energy.

To study the quarks confinement, we use the modified gluons propagation in deriving the static potential of the quark-quark gluons exchanging.\\
We define this potential in momentum space  using M matric element for quark-quark(gluons exchanging) interaction, with $\omega_0=k_0=0$(like Born approximation to the scattering amplitude in non-relativistic quantum mechanics[1])
\[
iM =  - i\tilde V(k)J^\mu  (p'_2 ,p_2 )J_\mu  (p'_1 ,p_1 )
\]
with the transferred current $J^\mu  (p',p) = \bar u(p')\gamma ^\mu  u(p)$ with spinor states $u(p)$ include the helicity states.\\

We find M matrix element using the Feynman diagrams for quark-quark gluons exchanging using color representation for one quark like
\[
u(p)_{color \otimes spinor}  = \frac{1}{{\sqrt 3 }}\left( \begin{array}{l}
 1 \\
 1 \\
 1 \\
 \end{array} \right)u(p)_{spinor}
\]
For distinguishable quarks(only one diagram), we have
\[
iM = \bar u^i (p'_2 )ig_s \gamma ^\mu  (T^a )_i ^j u_j (p_2 )\frac{{\Delta _{\mu \nu }^{ab} (k^2 )}}{i}\bar u^k (p'_1 )ig_s \gamma ^\nu  (T^b )_k ^\ell  u_\ell  (p_1 ){\rm{ }}
\]
with $k=p_2'-p_2 =p_1-p_1'$\\

Using Gell-Mann matrices, the matrices $T^a=\lambda^a; \lambda_1,\ldots, \lambda_8$ consider them as $SU(3)$ generators, using the modified gluons propagation we have
\[
iM = \sum\limits_{ijk\ell } {ig_s^2 } \bar u^i (p'_2 )\gamma ^\mu  (T^a )_i ^j u_j (p_2 )\frac{{g_{\mu \nu } \delta ^{ab} }}{{k^2 }}\left( {1 - \frac{{{\rm{a}}^{\rm{2}} k^2 }}{{1{\rm{ + a}}^{\rm{2}} k^2 }}} \right)\bar u^k (p'_1 )\gamma ^\nu  (T^b )_k ^\ell  u_\ell  (p_1 )
\]
to sum over the color indexes i, j with the color representation like above and over gluon index a we write
\[
\sum\limits_{ij} {\bar u^i } (p'_2 )\gamma ^\mu  (T^a )_i ^j u_j (p_2 ) = \bar u(p'_2 )\gamma ^\mu  \frac{1}{{\sqrt 3 }}\begin{pmatrix}
1 & 1 & 1\end{pmatrix}(T^a )\frac{1}{{\sqrt 3 }}\left( \begin{array}{l}
 1 \\
 1 \\
 1 \\
 \end{array} \right)u(p_2 )
\]
And $\frac{1}{{\sqrt 3 }}\begin{pmatrix}
1 & 1 & 1\end{pmatrix}(T^a )\frac{1}{{\sqrt 3 }}\left( \begin{array}{l}
 1 \\
 1 \\
 1 \\
 \end{array} \right) = \frac{1}{3}\sum\limits_{ij} {(T^a )_i ^j }$ \\
Therefore the M matrix element becomes
\[
M = \frac{1}{9}\sum\limits_a {\left( {\sum\limits_{ij} {(T^a )_i ^j } } \right)^2 } g_s^2 \bar u(p'_2 )\gamma ^\mu  u(p_2 )\frac{1}{{k^2 }}\left( {1 - \frac{{{\rm{a}}^{\rm{2}} k^2 }}{{1{\rm{ + a}}^{\rm{2}} k^2 }}} \right)\bar u(p'_1 )\gamma _\mu  u(p_1 )
\]
The Gell-Mann matrices with nonzero sum of the elements are
\[
\lambda _1  = \begin{pmatrix}
0 & 1 & 0\\
1 & 0 & 0\\
0 & 0 & 0\end{pmatrix} ,{\rm{ }}\lambda _4  =  \begin{pmatrix}
0 & 0 & 1\\
0 & 0 & 0\\
1 & 0 & 0\end{pmatrix}\text{and }   \lambda _6  = \begin{pmatrix}
0 & 0 & 0\\
0 & 0 & 1\\
0 & 1 & 0\end{pmatrix}
\]
So $\sum\limits_a {\left( {\sum\limits_{ij} {(T^a )_i ^j } } \right)^2 }  = 3 \cdot \left( 2 \right)^2  = 12$. \\
Therefore we have
\[
M = \frac{{12g_s^2 }}{9}\frac{1}{{k^2 }}\left( {1 - \frac{{{\rm{a}}^{\rm{2}} k^2 }}{{1{\rm{ + a}}^{\rm{2}} k^2 }}} \right)\bar u(p'_2 )\gamma ^\mu  u(p_2 )\bar u(p'_1 )\gamma _\mu  u(p_1 )
\]
we have the potential $\tilde V(k)$ in momentum space as we defined
\begin{align*}
iM &=  - i\tilde V(k)J^\mu  (p'_2 ,p_2 )J_\mu  (p'_1 ,p_1 )\\
&= i\frac{{12g_s^2 }}{9}g_s^2 \bar u(p'_2 )\gamma ^\mu  u(p_2 )\frac{1}{{k^2 }}\left( {1 - \frac{{k^2 }}{{k^2  + 1/{\rm{a}}^{\rm{2}} }}} \right)\bar u(p'_1 )\gamma _\mu  u(p_1 )
\end{align*}
With the transferred currents $J^\mu  (p'_2 ,p_2 ) = \bar u(p'_2 )\gamma ^\mu  u(p_2 )$ and $J^\mu  (p'_1 ,p_1 ) = \bar u(p'_1 )\gamma ^\mu  u(p_1 )$ \\
So we have
\[
\tilde V(k) =  - \frac{{4g_s^2 }}{3}\frac{1}{{k^2 }}\left( {1 - \frac{{k^2 }}{{k^2  + 1/{\rm{a}}^{\rm{2}} }}} \right)\]
Making Fourier transformation to the space XYZ, we have the static potential $U(x)$ $(k_0=0)$ like the electric potential[1]
\begin{align*}
U\left( x \right) &= \int {\frac{{d^3 k}}{{(2\pi )^3 }}\tilde V(k)} {\rm{ }}e^{ik \cdot x}  =  - \frac{{4g_s^2 }}{3}\int {\frac{{d^3 k}}{{(2\pi )^3 }}\frac{1}{{k^2 }}\left( {1 - \frac{{k^2 }}{{k^2  + 1/{\rm{a}}^{\rm{2}} }}} \right)} {\rm{ }}e^{ik \cdot x}\\
&=  - \frac{{4g_s^2 }}{{3 \cdot 4\pi r}}\left( {1 - \exp ( - \frac{r}{{\rm{a}}})} \right)\text{with }r = \sqrt {x^2  + y^2  + z^2 }
\end{align*}
For low limited energy we have $\rm{a}p>1$(the figure 2) so $r<\rm{a}$, the static potential becomes
\[
U\left( r \right) =  - \frac{{4g_s^2 }}{{3 \cdot 4\pi r}}\left[ {1 - \exp ( - \frac{r}{{\rm{a}}})} \right] = -u_0  + {\rm{a}}_{\rm{1}} r - {\rm{ a}}_{\rm{2}} r^{\rm{2}}  +  \ldots ..{\rm{ }}
\]
with
\[
u_{\rm{0}}  =   \frac{4}{3}\frac{{g_s^2 }}{{4\pi {\rm{a}}}} =  \frac{{4\alpha _s }}{{{\rm{3a}}}},\text{  }{\rm{a}}_1  = \sigma  = \frac{{g_s^2 }}{{3 \cdot 2\pi {\rm{a}}^2 }} = \frac{{2\alpha _s }}{{{\rm{3a}}^2 }},\text{  }{\rm{a}}_2  = \frac{{4\alpha _s }}{{3 \cdot 6{\rm{a}}^3 }}
\]
To fix $u_0= 4\alpha _s/3\rm{a}$ we write it like
\[u_0  =   \frac{{4\alpha _s }}{{{\rm{3a}}}} =   \frac{{4\alpha _s }}{{{\rm{3a}}^2 }}{\rm{a}} =  2\sigma {\rm{a}}\]
with fixing the string tension $\sigma$ and the length $\rm{a}\rightarrow\rm{a}_0$ at low energy. \\

This potential appears at low limited energy and prevents the quarks from spreading away, $r<\rm{a}$ so it holds the quarks inside the Hadrons. But the starting from the high energies $\rm{a}\rightarrow0$ , although the quarks masses are small but they are created only at high energies where they are free and by dropping the energy the situation $r<\rm{a}$ appears, the length a would run and becomes higher at low energies, so have ${-\rm{a}^2}k^2>1$ : $r<\rm{a}$ which is the confinement.\\
The confinement(at low limited energy) means when $\rm{r}\rightarrow\rm{a}$ the two interacted quarks kinetic energy becomes zero (ignore the quark mass), therefore the highest kinetic energy can the quark get equals $\sigma\rm{a}$ which relates to the potential $U(r)= -u_0+{\sigma}r+\ldots $ for $r<\rm{a}$ .\\

We can make $U(r)$ the potential for all quarks in $r<{\rm{a}}$ so $\sigma\rightarrow{\sum} {\sigma}$ and consider r as average distance between the interacted quarks, so the energy $\sigma {\rm{a}}$ becomes the highest kinetic energy of all quarks.\\
When $\rm{r}\rightarrow\rm{a}$ the potential becomes $U(0)=-u_0=-4{\alpha}_s/{\rm{3a}}=-\sigma {\rm{a}}<0$ therefore the total quarks energy becomes negative.\\

In this situation the free quarks disappear, they become condensed in the hadrons. So the role of the potential is reducing the number of the free quarks. Therefore the potential $u_0=\sigma {\rm{a}}$ lets to decrease the free quarks chemical potential ${\mu_0}$, we have
\[
{\rm{ }}\mu _0  \to \mu _0  + U(r) = \mu _0  - \frac{{\alpha _s }}{r}\left( {1 - e^{ - r/{\rm{a}}} } \right) = \mu (r) \approx \mu _0  - u_0  + \sigma r{\rm{ }}:{\rm{        }}r < {\rm{a}}
\]
we replaced $4{\alpha}_s/3$ with ${\alpha_s}$.\\
We renormalize this step at high energy for the free quarks, quarks plasma.

\section{The quarks Field dual behavior}
To have finite results in the perturbation interaction, we modified the propagation like
\begin{equation*}\numberwithin{equation}{section}
\overline \Delta  _{\mu \nu }^{ab} (k^2 ) = \frac{{g_{\mu \nu } \delta ^{ab} }}{{k^2  - i\varepsilon }}\left( {1 - \frac{{{\rm{a}}^{\rm{2}} k^2 }}{{1{\rm{ + a}}^{\rm{2}} k^2 }}} \right)\text{for glouns }
\end{equation*}
\begin{equation*}
\bar S_{ij} (\slashed{p}) = \frac{{ - \slashed{p}\delta _{ij} }}{{p^2  - i\varepsilon }}\left( {1 - \frac{{{\rm{a}}^2 p^2 }}{{1{\rm{ + a}}^2 p^2 }}} \right)\text{for quarks}
\end{equation*}
We saw we can ignore the modification terms $\rm{a}^2p^2 /(1+ \rm{a}^2p^2) $ and $ \rm{a}^2k^2 /(1+ \rm{a}^2k^2)$ at high energy, but when the energy drops down to limited energy, those terms take place, we can give them a physical meaning, for that we search for the corresponding terms in the Lagrange. \\

To do this, we find the role of those terms in the Feynman diagrams, in self energies, quarks gluons vertex,$\ldots$.\\
We find that the terms $\rm{a}^2p^2 /(1+ \rm{a}^2p^2) $ and $ \rm{a}^2k^2 /(1+ \rm{a}^2k^2)$ can be related to pairing quark-antiquark appears as scalar particles with mass $1/\rm{a}$ and charges $-1, 0, +1$ we can interrupt these particles as pions.\\

That appears in the particles-antiparticles composition in Feynman diagrams which mean for the fields, there is fields dual behavior, free fields and composite fields, this behavior lets to possibility for separating the particles and possibility for composition them, so the dual behavior of the fields is elementary behavior.\\
In general, for any particle $A$ and its antiparticle $\overline{A}$, in pertubation interaction, they pair and have a scalar particle $\overline{A}A$, this lets to reduce the currents(charges) of particles and antiparticle.

That is, for each outcoming particle, in Feynman diagrams, there is incoming antiparticle with positive energy and negative mass, depends on the coupling constant behavior (this is at high energy for the electromagnetic interaction and at low energy for the strong interaction, quarks and gluons). Therefore reducing their interactions with the charges in a way lets to finite results in the perturbation results.\\

Using the gluons modified propagation, the quark self-energy becomes(1.3)
\[
i\Sigma _{ij} (\slashed{p}) = 2g_s^2 C(R)\delta _{ij} \int {\frac{{d^4 \ell }}{{(2\pi )^4 }}} \frac{{( -\slashed{p}  - \slashed{\ell})}}{{(p + \ell )^2 }}\frac{1}{{\ell ^2 }}\left( {1 - \frac{{{\rm{a}}^{\rm{2}} \ell ^2 }}{{1{\rm{ + a}}^{\rm{2}} \ell ^2 }}} \right)
\]
we can separate it to two parts
\begin{flushleft}
1.	Quark$-$gluon part:
\end{flushleft}
\[
i\Sigma _{ij} (\slashed{p}) = 2g_s^2 C(R)\delta _{ij} \int {\frac{{d^4 \ell }}{{(2\pi )^4 }}} \frac{{( -\slashed{p} -\slashed{\ell} )}}{{(p + \ell )^2 }}\frac{1}{{\ell ^2 }}
\]
\begin{flushleft}
2.	pairing quarks part:
\end{flushleft}
\[
i\Sigma _{ij} (\slashed{p}) = 2g_s^2 C(R)\int {\frac{{d^4 \ell }}{{(2\pi )^4 }}} \frac{{( -\slashed{p} -\slashed{\ell}  )\delta _{ij} }}{{(p + \ell )^2 }}\frac{1}{{\ell ^2 }}\left( { - \frac{{{\rm{a}}^{\rm{2}} \ell ^2 }}{{1{\rm{ + a}}^{\rm{2}} \ell ^2 }}} \right)
\]
It appears in the pairing part there is a scalar field $\varphi$ propagation:
\[
\frac{1}{i}\frac{1}{{\ell ^2  + 1/{\rm{a}}^{\rm{2}} }}
\]
which is real scalar particles Field propagation with mass $1/{\rm{a}}$, to preserve the charges, spin,$\ldots$, this particle must be condensed of quark-antiquark $|\overline{q}q\rangle$ (particle-antiparticle in general) so we have new diagram(figure 3), we rewrite
\begin{figure}[h!]
  \includegraphics[width=0.55\textwidth]{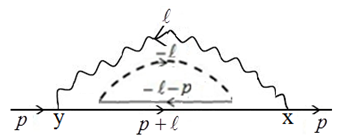}
  \caption{Representation the dual behavior, joined particle$-$antiparticle with opposite momentum$-$energy.}
\end{figure}
\[
i\Sigma _{ij} (\slashed{p}) = 2( - g_s )^2 C(R)\int {\frac{{d^4 \ell }}{{(2\pi )^4 }}} \frac{{( -\slashed{p} -\slashed{\ell} )\delta _{ij} }}{{i(p + \ell )^2 }}\frac{{ - i}}{{(\ell )^2  + 1/{\rm{a}}^{\rm{2}} }}
\]
Therefore we must add  new interaction terms to the quarks Lagrange, the possible terms are:
\[
\Delta L =  - ig_{\varphi q}\varphi \bar QQ\text{  }\text{ \text{ \text{ with } } }g_{\varphi q}  = g_s \sqrt {2C(R)}
\]
or
\[
\Delta L = g_{\varphi q} \varphi \bar Q\gamma _5 Q
\]
We expect the paring particles-antiparticles  preserves the flavor symmetry, so the real scalar field $\varphi$ becomes $|\overline{q_i}q_j\rangle$. For two flavors $q_i$ and $q_j$  we write the quarks field like $Q=\begin{pmatrix}
q_i & q_j\end{pmatrix}^T$ so
\[
\Delta L =  - ig_{\varphi q} \varphi ^a \bar QT_2^a Q{\text{  }}\text{ \text{ \text{ or } } }\Delta L = g_{\varphi q} \varphi ^a \bar QT_2^a \gamma _5 Q
\]
The real scalar fields $\varphi^a$ could interact with itself and have real non-zero ground value $\upsilon$ then $\langle\varphi\rangle=\upsilon$ so we can renormalize it like
\[
{\varphi^a}T_2^a\rightarrow\nu  - i\nu \pi ^a T_2^a  + ...
\]
then we have
\begin{align*}
\Delta L &=  - ig_{\varphi q} \bar Q(\nu  - i\nu \pi ^a T_2^a  + ...)Q \\
&=  - ig_{\varphi q} \nu \bar QQ - g_{\varphi \pi} \pi \bar QQ + ...;\text{Chiral symmetry breaking }
\end{align*}
or
\[
 \Delta L = g_{\varphi q} \bar Q(\nu  - i\nu \pi ^a T_2^a  + ...)\gamma _5 Q \to g_{\pi q} \bar Q\gamma _5 Q - ig_{\pi q} \pi \bar Q\gamma _5 Q + ...
\]
 here the particles $\pi ^a T_2^a  \to \pi  = (\pi ^0 ,\pi ^ -  ,\pi ^ + )$ are the pions.\\
The unusual terms $ - ig_{\varphi a} \nu \bar QQ\text{ and }g_{\pi q} \bar Q\gamma _5 Q$ are not hermitian and violate the symmetries, so they let the quarks disappear, damping at low energy $r<{\rm{a}}$ :
\begin{align*}
e^{i\Delta Et} \left| Q \right\rangle  &= e^{ - i\Delta Lt} \left| Q \right\rangle \\
&= e^{ - g_{\varphi q} \nu \bar qqt} \left| Q \right\rangle  = \sum\limits_n {e^{ - g_{\varphi q} \nu (\bar qq)t} } \left| {E_n } \right\rangle \left\langle {E_n } \right.\left| Q \right\rangle  \to \left| 0 \right\rangle \left\langle 0 \right.\left| Q \right\rangle
\end{align*}
$E_n$ is the energy of the quarks in state $\left| n \right\rangle$ and $e^{i\hat Ht} \left| Q \right\rangle$  is the eigenstate of the quarks field operator $\hat Q(t)$ in Heisenberg picture, $\hat Q(t) = e^{i\hat Ht} \hat Qe^{ - i\hat Ht}$.\\
\begin{figure}[h!]
  \includegraphics[width=0.55\textwidth]{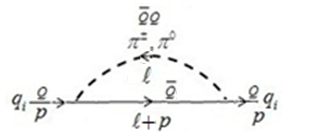}
  \caption{The quarks interaction with pions as a result of dual behavior.}
\end{figure}

That damping in the states is because of the pairing quark-antiquark at low energy ${\rm{a}}\neq0$, this pairing reduces the charges(currents) of free quarks(figure 5).\\
We can see that if we related the minus sign in $-{{{\rm{a}}^{\rm{2}} \ell ^2 }}/({{1{\rm{ + a}}^{\rm{2}} \ell ^2 }})$ to the fermions propagation:
\[
 S(x - y) = \int {\frac{{d^4 p}}{{(2\pi )^4 }}\frac{{ -\slashed{p}}}{{p^2 }}e^{ip(x - y)} }\text{ propagation from y to{\rm{ }}x }
\]
so
\[- S(x - y) =  - \int {\frac{{d^4 p}}{{(2\pi )^4 }}\frac{{ -\slashed{p} }}{{p^2 }}e^{ip(x - y)}  = \int {\frac{{d^4 p}}{{(2\pi )^4 }}\frac{{ +\slashed{p} }}{{p^2 }}e^{ip(x - y)} } }
\]
change $p\rightarrow -p$
\[- S(x - y) = \int {\frac{{d^4 p}}{{(2\pi )^4 }}\frac{{ -\slashed{p} }}{{p^2 }}e^{ - ip(x - y)}  = \int {\frac{{d^4 p}}{{(2\pi )^4 }}\frac{{ - \slashed{p} }}{{p^2 }}e^{ip(y - x)} } }\text{ propagation from x to y }
\]
So it is equivalent to invert the propagation $y{\rightarrow}x$ to $x{\rightarrow}y$ with positive energy and negative mass. Therefore it reduces the charges, currents, energies,\ldots of the particles and antiparticles,
\begin{figure}[h!]
  \includegraphics[width=0.55\textwidth]{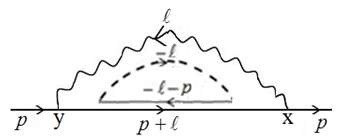}
  \caption{Omitting the distance x-y from the propagation.}
\end{figure}

we have(for one direction let it $y{\rightarrow}x$) 
\[
(p + \ell ) + ( - p - \ell ) = 0 \text{ and } (-\ell)+(\ell)=0\]
so incoming with p and outcoming with p, it is like to say the particles jump from y to x, in other words the distance y-x is removed from the interaction. \\
We expect the fields dual behavior takes place in negative potential. If there is no negative potential the paired particles would not survive(never condense).\\
For the quarks, the case $0<r<{\rm{a}}$ must associate with negative potential $u$ and $E+u<0$. Because the behavior of the strong interaction coupling constant at low energy, $\alpha_s$ is high, we expect negative potential at low energy $E+u<0 (E>0, u<0)$, so the quarks condense.\\

Because of the dual behavior of the quarks field which lets to quarks composite in scalar charged particles like the Pions ${\pi}^-, {\pi}^0, {\pi}^+$ and because of  their quantized charges ${-1, 0, +1}$ we expect the hadrons charges also quantized ${-}{Q}, - Q+1 ,\ldots, 0, +1,\ldots, +Q$  this quantization relates to the dual behavior of the quarks field in different hadrons, pairing quarks of different hadrons, so these condensed quarks; Pions, Kaons,\ldots are shared between the hadrons, so we put them together with the hadrons in groups, like the Pions ${-1, 0, +1}$ which can be inserted in SU(2) generators which can represent the proton-neutron pairing. \\

Therefore the protons and neutrons Lagrange contains the terms $- ig_{\pi N} \pi ^\alpha  \bar NT_2^\alpha  N$ with the nucleon field $N = \left( \begin{array}{l}
 p \\
 n \\
 \end{array} \right)
$.

\section{The Quarks Plasma}

We tried before to explain how the quarks are confined, for the strong interaction, we have the condition $r<\rm{a}\neq0$ at low limited energy and the condition $r>\rm{a}\rightarrow0$ at high energies for free quarks where the length a is removed from the propagators. But it appears to be fixed at low limited energy. In the last section we showed there is dual behavior for the quarks field, but when the length a is fixed, the result is scalar particles (pions) with mass $1/\rm{a}_0$ at low limited energy and the result is the chiral symmetry breaking.\\
We found the length a appears in the quark-quark strong interaction (gluons exchanging) potential $U(r)_{r<\rm{a}}<0$, so it relates to interaction strength. That is because, the behavior of the length a is like the behavior of the coupling constant $\alpha_s$ .\\
The confinement(at low energy $r<\rm{a}$) means when $r\rightarrow\rm{a}$ the two interacted quarks kinetic energy becomes zero (ignore the quark mass), therefore the highest kinetic energy can the quark get equals $\sigma\rm{a}$ which relates to the potential $U(r)=- u_0+\sigma r+\ldots $ for $r<\rm{a}$(at low limited energy).

When $r\rightarrow\rm{a}$ the potential becomes $U(0)=-u_0=-4\alpha_s/3\rm{a}<0$ therefore the total quarks energy becomes negative. In this situation the free quarks disappear$(\mu_{0}\rightarrow0)$, they become condensed in the hadrons.\\

We try here to use the statistical Thermodynamics to show how the free quarks disappear at low energies( low Temperatures) where the length a becomes fixed, so the chiral symmetry breaking and the quarks condensation.

One of the results is that the confinement phase(3.14) not necessary associates with chiral symmetry breaking, that is, the chiral symmetry breaking appears at the end of the cooling process when the expanding and cooling are ended and the length a becomes fixed, therefore the chiral symmetry breaking occurs and the pions become massive $m= 1/\rm{a}_0$.\\

We start with the massless quarks, their energy in volume V is
\begin{equation}
E = c\int\limits_{\rm{a}^3 } {d^3 r\int\limits_0^\infty  {d\varepsilon g(\varepsilon )\varepsilon \frac{1}{{e^{\beta \left( {\varepsilon  - \mu (r)} \right)}  + 1}}} } \text{ : }g(\varepsilon ) = g_q \frac{V}{{2\pi ^2 }}\varepsilon ^2
\end{equation}
\[
\text{ where }\mu (r) = \mu _0  + u(r)\text{ with }u(r) =  - \frac{{4\alpha _s }}{{3r}}\left( {1 - e^{ - r/{\rm{a}}} } \right)
\]
Here we inserted the quark-quark strong interaction potential $U(r)$ in the chemical potential (for decreasing the free quarks energy, as we think, the quarks potential reduces the free quarks chemical potential and make them condense at low energy) and because $r<\rm{a}$ we integrate over the volume $\rm{a}^3$: r is the distance between the interacted quarks. We can replace $4\alpha_s/3\rightarrow \alpha_s$ .\\

The constant c is determined by the comparing with free quarks high energy where the potential $U(r) \rightarrow0$ and $\alpha_s\rightarrow0$(decoupling) at high energies, so the length $\rm{a}\rightarrow0$ that is as we said before, the behavior of the length a is like the behavior of the coupling constant $g_s$  therefore the quarks become free at high energies.\\

By integrating over the energy(Maple program) we have:
\begin{align*}
E &= cg_q \frac{V}{{2\pi ^2 }}\int\limits_{{\rm{a}}^{\rm{3}} } {d^3 r\int\limits_0^\infty  {d\varepsilon \frac{{\varepsilon ^3 }}{{e^{\beta \left( {\varepsilon  - \mu (r)} \right)}  + 1}}} } \\
&= cg_q \frac{V}{{2\pi ^2 \beta ^4 }}\int\limits_{{\rm{a}}^{\rm{3}} } {d^3 r} \left[ {\frac{{7\pi ^4 }}{{60}} + \frac{{\pi ^2 }}{2}u_0 (r)^2  + \frac{1}{4}u_0 (r)^4  + 6\sum\limits_{k = 1}^\infty  {\frac{{( - 1)^k e^{ - k\beta \mu (r)} }}{{k^4 }}} } \right]
\end{align*}
with $u_0(r)=\beta\mu(r)=\beta(\mu_0+u(r))$ \\
by integrating over r (the distance between the interacted quarks) we have
\begin{eqnarray*}
   E&=& cg_q\frac{{2V{\rm{a}}^3 }}{{\pi x^4 }}\left[ {3.78 + 2(\beta \mu _0 )^2 \left( {0.82 - 1.16\frac{{\alpha _s }}{{{\rm{a}}\mu _0 }} + 0.41\left( {\frac{{\alpha _s }}{{{\rm{a}}\mu _0 }}} \right)^2 } \right)} \right.
\\
\\
    &&+ (\beta \mu _0 )^4 \left( {0.08 - 0.23\frac{{\alpha _s }}{{{\rm{a}}\mu _0 }} + 0.25\left( {\frac{{\alpha _s }}{{{\rm{a}}\mu _0 }}} \right)^2  - 0.12\left( {\frac{{\alpha _s }}{{{\rm{a}}\mu _0 }}} \right)^3  + 0.02\left( {\frac{{\alpha _s }}{{{\rm{a}}\mu _0 }}} \right)^4 } \right)
\\
\\
&& \text{ \text{ \text{ \text{ \text{ \text{ \text{ \text{ \text{ \text{ \text{ \text{ \text{ \text{ \text{ \text{ \text{ \text{ \text{ \text{ \text{ \text{ \text{ \text{  } } } } } } } } } } } } } } } } } } } } } } } }\text{  }\left. {{\rm{+ }}6\sum\limits_{k = 1}^\infty  {\int\limits_0^1 {x^2 dx} \frac{{( - 1)^k e^{ - k\beta \mu (x)} }}{{k^4 }}} } \right]
\label{eq:multilineeq}
\end{eqnarray*}

$g_q$ is the quarks degeneracy number and $x=\beta{\mu_0}$ .\\
For more easy we write $\alpha_s /{{\rm{a}}\mu_0} = 2\sigma{\rm{a}}/\mu_0 = y$ in the energy relation. So it becomes
\begin{eqnarray}
  E
   &=& cg_q\frac{{2V{\rm{a}}^3 }}{{\pi x^4 }}\left[ {3.78 + 2(\beta \mu _0 )^2 \left( {0.82 - 1.16y + 0.41y^2 } \right)} \right.\nonumber
\\ \nonumber
\\ \nonumber
    &&+ (\beta \mu _0 )^4 \left( {0.08 - 0.23y + 0.25y^2  - 0.12y^3  + 0.02y^4 } \right)\nonumber
\\
&& \text{ \text{ \text{ \text{ \text{ \text{ \text{ \text{ \text{ \text{ \text{ \text{  } } } } } } } } } } } }\text{  }\left. {{\rm{+ }}6\sum\limits_{k = 1}^\infty  {\int\limits_0^1 {x^2 dx} \frac{{( - 1)^k e^{ - k\beta \mu (x)} }}{{k^4 }}} } \right]
\label{eq:multilineeq}
\end{eqnarray}
at high energy: $x = \beta \mu _0  = \frac{{\mu _0 }}{T} \to 0$ \\
To find the constant c we compare with quarks high energy where they are free massless particles:
\[E_{high}  = g_q V\frac{{7\pi ^2 }}{{240}}T^4\]
When T is high, $x=(\mu_0 /T)\rightarrow0$  and $y\rightarrow0$ therefore $\beta\mu(x)\rightarrow0$ so we expand $e^{- k\beta \mu (x)}$ near $\beta\mu(x)=0$, we have:
\[E_{high}  = cg_q \frac{{2{\rm{a}}^{\rm{3}} V}}{{\pi x^4 }}\left[ {3.78 - 1.88 + O(x,y)} \right] \to cg_q \frac{{2{\rm{a}}^{\rm{3}} V}}{{\pi x^4 }} * 1.9
\]
\begin{equation}
\rightarrow g_q \frac{{7\pi ^2 V}}{{240}}T^4  = cg_q \frac{{2{\rm{a}}^{\rm{3}} V}}{{\pi x^4 }}1.9 \to c = \frac{\pi }{{2{\rm{a}}^{\rm{3}} 1.9}}\frac{{7\pi ^2 }}{{240}}\mu _0^4
\end{equation}
The energy becomes:
\begin{eqnarray*}
E
 &=&\frac{1}{{1.9}}\frac{{7\pi ^2 }}{{240}}\mu _0^4 g_q \frac{V}{{(\beta \mu _0 )^4 }} \left[ {3.78 + 2*(\beta \mu _0 )^2 \left( {0.82 - 1.16y + 0.41y^2 } \right)} \right.
\\
\\
    && \text{ \text{ \text{ \text{ \text{ \text{  } } } } } } + (\beta \mu _0 )^4 \left( {0.08 - 0.23y + 0.25y^2  - 0.12y^3  + 0.02y^4 } \right)
\\
\\
&& \text{ \text{ \text{ \text{ \text{ \text{ \text{ \text{ \text{ \text{ \text{ \text{  } } } } } } } } } } } }\text{ \text{ \text{ \text{ \text{ \text{  } } } } } }\left. {{\rm{+ }}6\sum\limits_{k = 1}^\infty  {\int\limits_0^1 {x^2 dx} \frac{{( - 1)^k e^{ - ku_0 (x)} }}{{k^4 }}} } \right]
\label{eq:multilineeq}
\end{eqnarray*}

Now we see the effects of the length a on the energy, at high energy, by fixing $x= \mu_0/T$ and varying $y=\sigma{\rm{a}}/2\mu_0 <1$:
\begin{eqnarray}
E_{high}
 &=& \frac{1}{{1.9}}\frac{{7\pi ^2 }}{{240}} g_q {V}{\mu _0^4{x^{-4} }} \left[ {1.9} \right. + x(1.8 - 1.24y) + x^2 (0.82 - 1.18y + 0.42y^2 )\nonumber
\\ \nonumber
\\
    && \text{ \text{ \text{ \text{ \text{ \text{ \text{ \text{ \text{ \text{ \text{ \text{  } } } } } } } } } } } }+ x^3 (0.23 - 0.47y + 0.33y^2  - 0.08y^3 )
\\ \nonumber
\\ \nonumber
&& \text{ \text{ \text{ \text{ \text{ \text{  } } } } } }+ x^4 (0.04 - 0.12y + 0.13y^2 \left. { - 0.07y^3  + 0.01y^4 ) +  \ldots } \right]_{x = \beta \mu _0  \to 0}\nonumber
\,.
\label{eq:multilineeq}
\end{eqnarray}
We expanded $e^{- k\beta \mu (x)}$ near $\beta\mu(x)=0$ and fixed the tension $\sigma$ as we assumed before, so we have the figure(6)
\begin{figure}[h!]
  \includegraphics[width=0.40\textwidth]{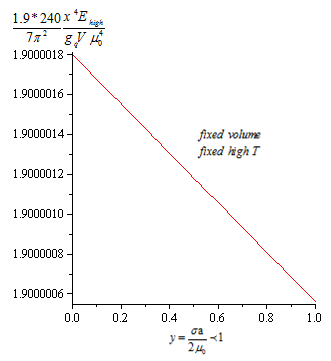}
  \caption{Decreasing the high energy with increasing y.}
\end{figure}

It appears in the figure that the high energy quarks lose an energy when the length a increases although the temperature is fixed. That means, when the length a increases the number of the excited quarks decreases.\\
That is because of the attractive linear potential  $\sigma r\ldots$  between the quarks, that potential absorbs an energy$( r<\rm{a}$ confinement, chapter 1), so the quarks are cooled faster by the expanding.
As we said before, the behavior of length a is like the behavior of the coupling constant $\alpha_s$ so when the energy dropped to lowest energy, the length a increased extremely and this is fast cooling(extremely cooling).\\
That occurs when the particles spread away, the length a, as a distance between the quarks, increases.\\

To determine the end, we search for the balance situations, such zero pressure, confinement condition,...\\
First we find the high energy pressure including the effects of the potential $\sigma \rm{a}$.\\
Starting from the general pressure relation:
\[
p =  - \frac{\partial }{{\partial V}}F\text{ where }F =  - T\ln Z =  - \frac{1}{\beta }\ln Z
\]
here we use  the relation:
\[
\ln Z = c\int\limits_{a^3 } {d^3 r\int\limits_0^\infty  {d\varepsilon g(\varepsilon )\ln \left( {e^{ - \beta \left( {\varepsilon  - \mu (r)} \right)}  + 1} \right)} } {\rm{  }}:{\rm{ }}g(\varepsilon ) = g_q \frac{V}{{2\pi ^2 }}\varepsilon ^2
\]
So the pressure becomes
\[
P = \frac{1}{3}\frac{\partial }{{\partial V}}E
\]
so for high energy $x = \beta \mu _0  \to 0$ we have the pressure:
\begin{eqnarray*}
P_{high}
 &=& \frac{1}{3}\frac{\partial }{{\partial V}}E_{high}= \frac{\partial }{{\partial V}}\frac{1}{{3 \cdot 1.9}}\frac{{7\pi ^2 }}{{240}}g_q V\mu _0^4 x^{ - 4} \left[ {1.9} \right. + x(1.8 - 1.24y)
\\
\\
    && \text{  }+ x^2 (0.82 - 1.18y + 0.42y^2 )+ x^3 (0.23 - 0.47y + 0.33y^2  - 0.08y^3 )
\\ \nonumber
\\ \nonumber
&& \text{ \text{ \text{ \text{ \text{ \text{  } } } } } }+ x^4 (0.04 - 0.12y + 0.13y^2 \left. { - 0.07y^3  + 0.01y^4 ) +  \ldots } \right]\nonumber
\label{eq:multilineeq}
\end{eqnarray*}

Now the key point is, we want to include the potential effect on the pressure so we replace the volume V with the volume $\rm{a}^3 \sim y^3$ so
\\
\begin{eqnarray}
P_{high}
 &\rightarrow& \frac{\partial }{{\partial y^3 }}y^3 \frac{1}{{3 \cdot 1.9}}\frac{{7\pi ^2 }}{{240}}g_q \mu _0^4 x^{ - 4} \left[ {1.9} \right. + x(1.8 - 1.24y)
\\ \nonumber
\\ \nonumber
    && \text{  }+ x^2 (0.82 - 1.18y + 0.42y^2 )+ x^3 (0.23 - 0.47y + 0.33y^2  - 0.08y^3 )
\\ \nonumber
\\ \nonumber
&& \text{ \text{ \text{ \text{ \text{ \text{  } } } } } }+ x^4 (0.04 - 0.12y + 0.13y^2 \left. { - 0.07y^3  + 0.01y^4 ) +  \ldots } \right]\nonumber
\label{eq:multilineeq}
\end{eqnarray}
Which is represented in the figure(7), without conditions on y or on the length a
\begin{figure}[h!]
  \includegraphics[width=0.40\textwidth]{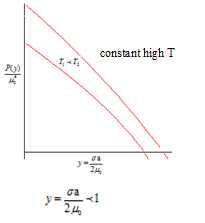}
  \caption{The effects of potential $\sigma\rm{a}$ on the pressure.}
\end{figure}

It is clear(without conditions on y) the pressure decreases with increasing the length a (decreasing the quarks energy $-p^2$)  until it becomes zero, then negative.\\
That becomes clear at low energy where there are conditions on y and so on the length a.\\

For the low energy quarks, $T\rightarrow0$ so ${\beta}\mu(x)\rightarrow \infty$  so  $e^{ - k\beta \mu (x)}\rightarrow0$. The energy becomes:
\begin{eqnarray}
E_{low}
 &=& \frac{1}{{1.9}}\frac{{7\pi ^2 }}{{240}}\mu _0^4 g_q \frac{V}{{(\beta \mu _0 )^4 }}
\left[ {3.78 + 2*(\beta \mu _0 )^2 \left( {0.82 - 1.16y + 0.41y^2 } \right)} \right.\nonumber \\
\nonumber  \\
    && \text{  } {\rm{ }}\left. { + (\beta \mu _0 )^4 \left( {0.08 - 0.23y + 0.25y^2  - 0.12y^3  + 0.02y^4 } \right)} \right]{\rm{ }}
\label{eq:multilineeq}
\end{eqnarray}
Making $x=T/\mu_0$ so
\begin{eqnarray*}
E_{low}
 &=& \frac{1}{{1.9}}\frac{{7\pi ^2 }}{{240}}\mu _0^4 g_q{V}{{x^4 }}\left[ {3.78 + 2*x^{-2} \left( {0.82 - 1.16y + 0.41y^2 } \right)} \right.\nonumber \\
\nonumber  \\
    &&\text{ \text{ \text{ \text{ \text{  } } } } } {\rm{ }}\text{  } {\rm{ }} \text{  } {\rm{ }}\left. { + x^{-4} \left( {0.08 - 0.23y + 0.25y^2  - 0.12y^3  + 0.02y^4 } \right)} \right]{\rm{ }}
\label{eq:multilineeq}
\end{eqnarray*}
Now the key point, we want to show the effect of the potential $\sigma{\rm{ a}}$ on the energy so we see the behavior of the energy in the volume $\rm{ a}^3$ with respect to $y=2\sigma{\rm{ a}}/\mu_0$ the diagram is
\begin{figure}[h!]
  \includegraphics[width=0.40\textwidth]{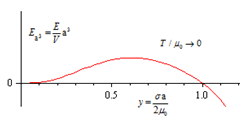}
  \caption{The extremely decreasing in quarks low energy in the strong interaction.}
\end{figure}

That is extremely behavior after $y=0.6$ where the energy $(E/V){\rm{ a}}^3$ decreases when the volume $\rm{ a}^3$ increases, the end in $y=1$ where the free quarks disappear for $y>1$ \\

Now we can distinguish between the confinement and the chiral symmetry breaking, when $y>0.6$ there is confinement: extremely cooling, negative pressure.
But when reach $y=1$ there is chiral symmetry breaking where the length a becomes fixed, and from the quarks  field dual behavior  there are scalar charged particles with mass $1/{\rm{ a}}$ appear when the length a is fixed with non-zero value ${\rm{ a}}_0$ .\\
Here the evidence for fixing the length a is the lowest limited quarks energy, that is as we said before, the behavior of the length a is like the behavior of the coupling constant $\alpha_s$ so when the quarks energy dropped (extremely cooling) the length a increases extremely to reach the highest value when $y=1$ which equivalents to smallest energy $E=0$ (the cooling end).\\
Another evidence for fixing the length a (chiral symmetry breaking) is the low energy pressure:
\[
P_{low}  = \frac{1}{3}\frac{\partial }{{\partial V}}E_{low}  \to \frac{1}{3}\frac{\partial }{{\partial y^3 }}\frac{{E_{low} }}{V}y^3
\]
To include the potential effect we study the pressure using the volume $\rm{a}^3\sim y^3$ therefore
\begin{eqnarray*}
  P_{low} &\rightarrow& \frac{1}{3}\frac{\partial }{{\partial y^3 }} \frac{1}{{1.9}}\frac{{7\pi ^2 }}{{240}}g_q\mu _0^4  y^3 x^4 \left[ {3.78 + 2x^{ - 2} \left( {0.82 - 1.16y + 0.41y^2 } \right)} \right.
\nonumber
\\
   &&   \text{ \text{ \text{ \text{ \text{ \text{  } } } } } }\left. { + x^{ - 4} \left( {0.08 - 0.23y + 0.25y^2  - 0.12y^3  + 0.02y^4 } \right)} \right]
\,.
\label{eq:multilineeq}
\end{eqnarray*}
therefore
\begin{eqnarray}
 && \frac{{P_{low} }}{{\mu _0^4 }}= \frac{1}{{9*1.9}}\frac{{7\pi ^2 }}{{240}}g_q \left[ {3 \cdot 3.78x^4  + 3*2*x^2 \left( {0.82 - 1.16y + 0.41y^2 } \right)} \right. \nonumber\\
\nonumber\\
   &&   \text{  }\text{  }{\rm{ + 3*(}}0.08 - 0.23y + 0.25y^2  - 0.12y^3  + 0.02y^4 ){\rm{ + }}2yx^2 \left( { - 1.16 + 0.82y} \right) \nonumber\\
   \nonumber\\
  && \text{ \text{ \text{ \text{ \text{ \text{ \text{ \text{ \text{ \text{ \text{ \text{  } } } } } } } } } } } }\text{  }\left. { + y\left( { - 0.23 + 0.5y - 0.36y^2  + 0.08y^3 } \right)} \right]
\label{eq:multilineeq}
\end{eqnarray}
we see its behavior in figure(9)bellow
\begin{figure}[h!]
  \includegraphics[width=0.40\textwidth]{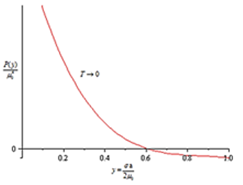}
  \caption{The extremely decreasing in the pressure at low energy.}
\end{figure}

it is clear from the figure, when $y>0.6$ the quarks pressure becomes negative.\\
We expect the condensed quarks phase (confinement quarks) has positive pressure, so the preferred phase is the condensed quarks phase.\\
So when $y>0.6$ the quarks condense until $y=1: \rm{a}\rightarrow\rm{a}_0\approx1/(135-140Mev)$ the quarks disappear, the scalar charged particles(Pions) appear instead of them, that is because of the quarks dual behavior(free-condensed quarks), but at low limited energy the condensed phase has a big chance instead the free Phase.\\

\emph{The Confinement phase}:\\
In this paper we study two quarks (up and down) condensation in the pions $(\pi^0, \pi^+, \pi^-)$ and baryons$(n, p^+, p^-)$, so the degeneracy number is $g_q=2_{flavor}\ast2_{charge}\ast2_{spin}\ast3_{color}=24$ .\\

We need more clarifying for determining if the quarks could stay free particles or they condense in hadrons.\\
We can think they could be free if their energy is enough for covering the strong interaction potential and stay free particles with least possible energy(at 0 temperature). Unless they condense in the hadrons.

To cover the strong interaction potential means to lose an energy $E_u$ which is transferred to the exchanged static gluons and pions which are created between the low energy quarks. So the remains energy in the volume $4\pi \rm{a}^3/3$  is
\begin{equation}
\frac{{E_{q,low} }}{V}\frac{{4\pi }}{3}{\rm{a}}^3  - \frac{{E_u }}{V}\frac{{4\pi }}{3}{\rm{a}}^3
\end{equation}
This energy must be enough for the least possible free quarks. Therefore we must determine the chemical potential $\mu_0$ of the free quarks with smallest possible density at 0 Temperature.

According to the quarks confinement $r<\rm{a}$ at low limited energy, which means the highest possible distance between the two interacted quarks is a, we expect the least quarks density is two quarks in the volume $4\pi (\rm{a/2})^3/3$ .
\begin{figure}[h!]
  \includegraphics[width=0.45\textwidth]{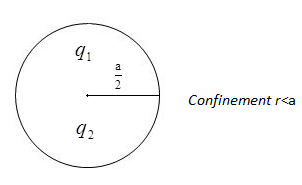}
  \caption{The quarks confinement at low energy.}
\end{figure}

From this view we can calculate the least quarks chemical potential $\mu_0$ of free quarks:
\[
\frac{2}{{\frac{4\pi}{3} ({\rm{a/2)}}^3}} = \frac{1}{V}\int\limits_0^{\mu _0 } {g(\varepsilon )d\varepsilon  = g_q \frac{{\mu _0^3 }}{{6\pi ^2 }}}  \to (\mu _0 {\rm{a/2)}}^3  = \frac{{9\pi }}{{g_q }} \to (\mu _0 {\rm{a)}}^3  = \frac{{8*9\pi }}{{g_q }}
\]
$1/{\rm{a}}$ is the pion mass when ${\rm{a}}\rightarrow{\rm{a}}_0$ in the end of free quarks phase so $1/{\rm{a}}\rightarrow(135-140)Mev$.
So the least free quarks energy density in 0 temperature is
\[
\frac{{\varepsilon _{free} {\rm{ }}}}{V} = \frac{1}{V}\int\limits_0^{\mu _0 } {g(\varepsilon )\varepsilon d\varepsilon  = g_q \frac{{\mu _0^4 }}{{4 \cdot 2\pi ^2 }}}
\]
So the smallest energy of the free quarks in the volume $4\pi{\rm{a}}^3/3$  is
\begin{align}
\varepsilon _{free,{\rm{ a}}^3 }  = g_q \frac{{\mu _0^4 }}{{4 \cdot 2\pi ^2 }}\frac{{4\pi {\rm{a}}^3 }}{3}&= \frac{{4\pi }}{3}g_q \frac{{\mu _0 }}{{4 \cdot 2\pi ^2 }}{\rm{(}}\mu _0 {\rm{a)}}^3   \nonumber\\
\nonumber\\
&= \frac{{4\pi }}{3}g_q \frac{{\mu _0 }}{{4 \cdot 2\pi ^2 }}\frac{{8*9\pi }}{{g_q }} = \frac{{4\pi }}{3}\frac{9}{\pi }\mu _0
\end{align}
therefore
\[
\frac{{\varepsilon _{free,{\rm{ a}}^3 } }}{{2\mu _0 }} = \frac{{4\pi }}{3}\frac{9}{{2\pi }} = \frac{{4\pi }}{3}*{\rm{1}}{\rm{.43}}
\]
Because the chemical potential $\mu\sim1/{\rm{a}}$ and $\mu\rightarrow\mu_0$ when $\rm{a}\rightarrow\rm{a}_0$ and because $y\sim \rm{a}$ so we modified  $\mu_0\rightarrow \mu_0/y$ so
\begin{equation}
\frac{{4\pi }}{3}*{\rm{1}}{\rm{.43}} \to \frac{{4\pi }}{3}*\frac{{{\rm{1}}{\rm{.43}}}}{y}{\rm{ }}
\end{equation}
Now we find the least energy $E_u$ which is transferred to the static exchanged gluons and pions according to the potential
\[
{\rm{ }}u(r) =  - \frac{{4\alpha _s }}{{3r}}\left( {1 - e^{ - r/{\rm{a}}} } \right) \approx  - u_0  + \sigma r{\rm{ }}:{\rm{   }}r < {\rm{a}}
\]
We absorbed $4/3$ to $\alpha _s$ so   and made $\alpha _s /{\rm{a}\mu_0} = 2\sigma\rm{a}/\mu_0 = y$ the constant $\sigma$ is the string tension.\\
This potential is inserted to reduce the chemical potential $\mu_0$ and the energy is renormalized at high energy. So we have $\mu_0\rightarrow \mu_0 + u(r)$ :
\[
{\rm{ }}\mu (r) = \mu _0  - \frac{{\alpha _s }}{r}\left( {1 - e^{ - r/{\rm{a}}} } \right) \approx \mu _0  - u_0  + \sigma r{\rm{ }}:{\rm{   }}r < {\rm{a}}
\]
Therefore we can calculate the least absorbed energy by this potential, by calculating the changes on the energy density at 0 temperature
\begin{align*}
\frac{{\varepsilon (\alpha _s /{\rm{a}})}}{V} &= \frac{{\varepsilon (y)}}{V}= \frac{1}{V}c\int\limits_0^{\rm{a}} {4\pi r^2 dr} \int\limits_0^{\mu (r)} {g(\varepsilon )\varepsilon d\varepsilon }  \\
&= c\int\limits_0^{\rm{a}} {4\pi r^2 dr} g_q \frac{{\mu (r)^4 }}{{4 \cdot 2\pi ^2 }}
\end{align*}
The constant c is determined $c = \frac{\pi }{{2{\rm{a}}^{\rm{3}} 1.9}}\frac{{7\pi ^2 }}{{240}}$ so the interaction energy is
\begin{align*}
\frac{{\varepsilon (\alpha _s /{\rm{a}})}}{V} = \frac{{\varepsilon (y)}}{V} &= g_q \frac{\pi }{{2{\rm{a}}^{\rm{3}} 1.9}}\frac{{7\pi ^2 }}{{240}}\frac{{4\pi }}{{4 \cdot 2\pi ^2 }}\int\limits_0^{\rm{a}} {r^2 dr} \mu (r)^4 \\
& = g_q \frac{{7\pi ^2 }}{{4*1.9*240{\rm{a}}^{\rm{3}} }}\int\limits_0^{\rm{a}} {r^2 dr} \mu (r)^4
\end{align*}
This becomes
\begin{align*}
\frac{{\varepsilon (\alpha _s /{\rm{a}})}}{V} = \frac{{\varepsilon (y)}}{V}&= g_q \frac{{7\pi ^2 }}{{4*1.9*240{\rm{a}}^{\rm{3}} }}\int\limits_0^{\rm{a}} {r^2 dr} \left[ {\mu _0  - \frac{{\alpha _s }}{r}\left( {1 - e^{ - r/a} } \right)} \right]^4  \\
&= g_q \frac{{7\pi ^2 }}{{4*1.9*240{\rm{a}}^{\rm{3}} }}(\mu _0 )^4 \int\limits_0^{\rm{a}} {r^2 dr} \left[ {1 - \frac{{\alpha _s }}{{\mu _0 r}}\left( {1 - e^{ - r/a} } \right)} \right]^4
\end{align*}
Using the changing $r=\rm{a}x$ so
\[
\frac{{\varepsilon (\alpha _s /{\rm{a}})}}{V} = \frac{{\varepsilon (y)}}{V} = g_q \frac{{7\pi ^2 }}{{4*1.9*240{\rm{a}}^{\rm{3}} }}(\mu _0 )^4 \int\limits_0^{\rm{1}} {{\rm{a}}^3 x^2 dx} \left[ {1 - \frac{{\alpha _s }}{{\mu _0 {\rm{a}}x}}\left( {1 - e^{ - x} } \right)} \right]^4
\]
therefore
\[
\frac{{\varepsilon (\alpha _s /{\rm{a}})}}{V} = \frac{{\varepsilon (y)}}{V} = g_q \frac{{7\pi ^2 }}{{4*1.9*240}}(\mu _0 )^4 \int\limits_0^{\rm{1}} {x^2 dx} \left[ {1 - \frac{y}{x}\left( {1 - e^{ - x} } \right)} \right]^4
\]
The spent energy for the interaction in the volume $4\pi \rm{a}^3/3$ is
\begin{align}
\varepsilon _{u,{\rm{a}}^3 }  &= \frac{{\varepsilon (1) - \varepsilon (0)}}{V}\frac{{4\pi {\rm{a}}^3 }}{3}  \\
\nonumber\\
&= \frac{{4\pi }}{3}g_q \frac{{7\pi ^2(\mu _0 {\rm{a}})^3\mu _0 }}{{4*1.9*240}}  \left( {\int\limits_0^{\rm{1}} {x^2 dx} \left[ {1 - \frac{1}{x}\left( {1 - e^{ - x} } \right)} \right]^4  - \int\limits_0^{\rm{1}} {x^2 dx} } \right)\nonumber
\end{align}
it becomes
\begin{align*}
\varepsilon _{u,a^3 }= \frac{{\varepsilon (1) - \varepsilon (0)}}{V}\frac{{4\pi {\rm{a}}^3 }}{3} &=  - \frac{{4\pi }}{3}g_q \frac{{7\pi ^2 }}{{4*1.9*240}}(\mu _0 {\rm{a}})^3 \mu _0 0.33 \\
\\
&=  - \frac{{4\pi }}{3}g_q \frac{{7\pi ^2 }}{{4*1.9*240}}\frac{{8*9\pi }}{{g_q }}\mu _0 \ast0.33
\end{align*}
Therefore
\begin{align}
\varepsilon _{u,{\rm{a}}^3 }  = \frac{{\varepsilon (1) - \varepsilon (0)}}{V}\frac{{4\pi {\rm{a}}^3 }}{3}& =  - \frac{{4\pi }}{3}g_q \frac{{7\pi ^2 }}{{4*1.9*240}}(\mu _0 {\rm{a}})^3 \mu _0  * 0.33  \\
 \nonumber\\
& =  - \frac{{4\pi }}{3}\frac{{7*8*9*0.33\pi ^3 }}{{4*1.9*240}}\mu _0  =  - \frac{{4\pi }}{3}{\rm{*2}}{\rm{.82}}\mu _0 \nonumber
\end{align}
So we have
\[
\frac{{\varepsilon _{u,{\rm{a}}^3 } }}{{2\mu _0 }} =  - \frac{{4\pi }}{3}{\rm{*1}}{\rm{.41}}
\]
As for $E_{free}$ we replace
\[
\frac{{4\pi }}{3}{\rm{*1}}{\rm{.41}} \to \frac{{4\pi }}{3}{\rm{*}}\frac{{{\rm{1}}{\rm{.41}}}}{y}
\]
 Now we find the confinement condition at any temperature, if the quarks energy is not enough to cover the interaction energy $E_u$ and give free quarks with smallest density, at 0 temperature, then they become confinement$(r<{\rm{a}})$, so the confinement condition
\begin{equation}
E(T,y) - \varepsilon _u  - \varepsilon _{free}  \prec 0
\end{equation}
Then
\[
\frac{{E(T,y)}}{V}\frac{{4\pi {\rm{a}}^3 }}{3} - \frac{{\varepsilon _u }}{V}\frac{{4\pi {\rm{a}}^3 }}{3} - \frac{{\varepsilon _{free} }}{V}\frac{{4\pi {\rm{a}}^3 }}{3} \prec 0
\]
Or
\[
\frac{{E(T,y)}}{{2\mu _0 V}}\frac{{4\pi {\rm{a}}^3 }}{3} - \frac{{\varepsilon _u }}{{2\mu _0 V}}\frac{{4\pi {\rm{a}}^3 }}{3} - \frac{{\varepsilon _{free} }}{{2\mu _0 V}}\frac{{4\pi {\rm{a}}^3 }}{3} \prec 0
\]
We consider
 \[\sigma _{{\rm{a}}^3 } {\rm{a = }}\frac{{\varepsilon _{free} }}{V}\frac{{4\pi {\rm{a}}^3 }}{3}\]
as critical energy of free quarks for lowest energy, the tension $\sigma_{\rm{a}^3}$ here is the volume tension.\\
Therefore this critical energy is transferred to the produced  hadrons and photons.\\
Using the quarks low energy
\begin{eqnarray*}
  E_{low} &=& \frac{1}{{1.9}}\frac{{7\pi ^2 }}{{240}}g_q \mu _0^4 Vx^4 \left[ {3.78 + 2*x^{ - 2} \left( {0.82 - 1.16y + 0.41y^2 } \right)} \right.
\\
\\
   &&    \text{ \text{ \text{ \text{ \text{ \text{  } } } } } }\left. { + x^{ - 4} \left( {0.08 - 0.23y + 0.25y^2  - 0.12y^3  + 0.02y^4 } \right)} \right]
\label{eq:multilineeq}
\end{eqnarray*}

With  $x=T/{\mu_0}<<1$\\

The confinement condition becomes
\begin{eqnarray*}
  && \frac{1}{2}\frac{1}{{1.9}}\frac{{7\pi ^2 }}{{240}}\mu _0^3 g_q x^4 \frac{{4\pi {\rm{a}}^3 }}{3}\left[ {3.78 + 2*x^{ - 2} \left( {0.82 - 1.16y + 0.41y^2 } \right) + x^{ - 4} \left( {0.08} \right.} \right.{\rm{ }}
\\
\\
   && \text{ \text{ \text{ \text{ \text{ \text{ \text{ \text{ \text{  } } } } } } } } } \left. {\left. { - 0.23y + 0.25y^2  - 0.12y^3  + 0.02y^4 } \right)} \right] - \frac{{\varepsilon _u }}{{2\mu _0 V}}\frac{{4\pi {\rm{a}}^3 }}{3} - \frac{{\varepsilon _{free} }}{{2\mu _0 V}}\frac{{4\pi {\rm{a}}^3 }}{3} \prec 0{\rm{ }}
\label{eq:multilineeq}
\end{eqnarray*}
It becomes
\begin{eqnarray*}
  & & \frac{1}{2}\frac{1}{{1.9}}\frac{{7\pi ^2 }}{{240}}g_q \frac{{4\pi (\mu _0 {\rm{a)}}^3 }}{3}\left[ {3.78x^4  + 2*x^2 \left( {0.82 - 1.16y + 0.41y^2 } \right)} \right.
\\
\\
   && \text{ \text{ \text{ \text{ \text{ \text{  } } } } } } \left. { + \left( {0.08 - 0.23y + 0.25y^2  - 0.12y^3  + 0.02y^4 } \right)} \right] - \frac{{4\pi }}{3}{\rm{*}}\frac{{{\rm{1}}{\rm{.41}}}}{y} - \frac{{4\pi }}{3}*\frac{{{\rm{1}}{\rm{.43}}}}{y} \prec 0{\rm{ }}
\label{eq:multilineeq}
\end{eqnarray*}
We had the relation
 \[
 (\mu _0 {\rm{a)}}^3  = \frac{{8*9\pi }}{{g_q }}
 \]
therefore, the condition becomes
\begin{eqnarray*}
  & & \frac{1}{2}\frac{1}{{1.9}}\frac{{7\pi ^2 }}{{240}}\frac{{4*8*9\pi ^2 }}{3} \left[ {3.78x^4  + 2*x^2 \left( {0.82 - 1.16y + 0.41y^2 } \right)} \right.
\\
\\
   && \text{ \text{ \text{ \text{ \text{ \text{  } } } } } } \left. { + \left( {0.08 - 0.23y + 0.25y^2  - 0.12y^3  + 0.02y^4 } \right)} \right] - \frac{{4\pi }}{3}{\rm{*}}\frac{{{\rm{1}}{\rm{.41}}}}{y} - \frac{{4\pi }}{3}*\frac{{{\rm{1}}{\rm{.43}}}}{y} \prec 0{\rm{ }}
\label{eq:multilineeq}
\end{eqnarray*}
it becomes
\begin{equation}
\begin{array}{l}
 3.78x^4  + 2*x^2 \left( {0.82 - 1.16y + 0.41y^2 } \right) \\
 \\
  \text{ \text{ \text{  } } }+ \left( {0.08 - 0.23y + 0.25y^2  - 0.12y^3  + 0.02y^4 } \right) - 0.16{y^{-1}} \prec 0 \\
 \end{array}
\end{equation}
with the curve
\begin{figure}[h!]
  \includegraphics[width=0.45\textwidth]{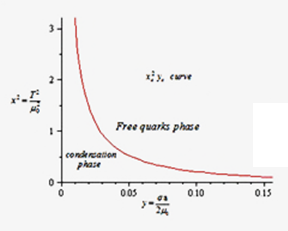}
  \caption{The critical $X_c ^2Y_c$ curve separates the free and confinement quarks phases.}
\end{figure}

The critical situation $x_c$ with $y\rightarrow1$ (the end of the extremely cooling)
\[
3.78x_c ^4  + 2*{\rm{0}}{\rm{.07}}x_c ^2  - {\rm{ 0}}{\rm{.16574 = }}0{\rm{ }} \to {\rm{ }}x_c  = {\rm{0}}{\rm{.438}}
\]

So the critical temperature of the confinement condition when $y\rightarrow1$  from  $x_c=T_c/{\mu_0}$ is $T_c=0.438{\mu_0}$ \\

We determine ${\mu_0}$ from
\[
 (\mu _0 {\rm{a)}}^3  = \frac{{8*9\pi }}{{g_q }}
\]
when $y\rightarrow1$ so $\rm{a}\rightarrow\rm{a}_0$ we set $1/\rm{a}_0=$ pion mass $=(135-140)Mev$ so
\[
\mu _0  = \frac{1}{{\rm{a}}}\left( {\frac{{8*9\pi }}{{g_q }}} \right)^{\frac{1}{3}}  \to 135*\left( {\frac{{8*9\pi }}{{24}}} \right)^{\frac{1}{3}}  = {\rm{285}}{\rm{.15}}Mev{\rm{  }}\text{ for }{\rm{  }}\frac{1}{{{\rm{a}}_0 }} = 135Mev
\]
So the critical temperature is $T_c=0.438*285.15=124.9Mev$.\\

Now we try to find the produced Hadrons, after covering the potential (3.12), the quarks critical energy(possible smallest energy) $E_{free}$ (3.9) is transferred to the produced hadrons and photons.\\

The key Idea here is: because the cooling is an extremely cooling, it is expanding  $\rm{a}:0\rightarrow\rm{a}_0=1/(135-140Mev)$ so this process is thermally isolated from the
other fields(adiabatic changing), therefore the produced particles are in $Tc=124.9Mev$.\\
We assume that the produced particles are hadrons(fermions and bosons) and photons.\\
When $\rm{a}:0\rightarrow\rm{a}_0 : y\rightarrow1$ the pions become massive $m=1/\rm{a}_0$ so we expect the other hadrons become massive at this stage, we assume that is in $T\rightarrow T_c$.\\

Therefore we assume when $T>T_c$ massless hadrons and $T<T_c$  massive hadrons. Anyway in $x_cy_c$ curve we find the confinement is possible at high energy$(T>>T_c : \rm{a}\rightarrow0)$.\\
First we write using (3.9)
\begin{align}
\frac{{\varepsilon _{free} }}{V} = \frac{{\varepsilon _{free,{\rm{ a}}^3 } }}{{4\pi {\rm{a}}^3 /3}} &= g_q \frac{{\mu _0^4 }}{{4 \cdot 2\pi ^2 }} \\
&= \frac{{\sigma _{{\rm{a}}^3 } {\rm{a}}}}{{4\pi {\rm{a}}^3 /3}} \to \frac{{E_{hadrons}  + E_{photons} }}{V}{\rm{ }}\text{ below }x_c y_c \text{ curve } \nonumber
\end{align}
or
\[
\frac{{\sigma _{{\rm{a}}^3 } {\rm{a}}}}{{4\pi {\rm{a}}^3 /3}} = g_q \frac{{\mu _0^4 }}{{4 \cdot 2\pi ^2 }} \to \varepsilon _f  + \varepsilon _b  + \varepsilon _{ph}
\]
With the densities
\[
\varepsilon _f  = \frac{{E_f }}{V}{\rm{ }}\text{ , }\varepsilon _b  = \frac{{E_b }}{V}{\rm{ }}\text{ and }{\rm{ }}\varepsilon _{ph}  = \frac{{E_{ph} }}{V}
\]
for spin $1/2$ hadrons $(fermions \text{   }p^+, p^-, n)$, spin 0 hadrons $(bosons \text{   }\pi^0, \pi^-, \pi^+)$ and photons densities.\\
For massless phase $T>>T_c$ and $y_c\approx0$ ignoring the chemical potential we have
\begin{equation}
n_f  = \frac{{N_f }}{V} = g_f \frac{{3\zeta (3)}}{{4\pi ^2 }}T^3 {\rm{  }},{\rm{  }}n_b  = \frac{{N_b }}{V} = g_b \frac{{\zeta (3)}}{{\pi ^2 }}T^3 {\rm{ }}\text{ and }{\rm{ }}\varepsilon _{ph}  = \frac{{E_{ph} }}{V} = g_{ph} \frac{{\pi ^2 }}{{30}}T^4 {\rm{ }}
\end{equation}
Now the key point, because the cooling is extremely cooling, like to take all the particles (quarks) from high Temperature and put them at low Temperature, so the same structure at high energy will be at low energies, like the charges ratios, energy distribution over the particles, spins,\ldots \\

At $T\rightarrow T_c$ and $y_c=1$ the hadrons become massive, we approximate:\\
for bosons(pions with mass $1/{\rm{a }}_0=135-140Mev$) the energy density becomes:
\begin{align*}
 \varepsilon _b  &= g_b \frac{{\pi ^2 }}{{30}}T^4  \to \varepsilon _b  = g_b \frac{{\pi ^2 }}{{30}}T^4  + m_{pion} n_b {\rm{  }} \\
 \\
 &\text{ with }{\rm{  }}n_b  = \frac{{N_b }}{V} = g_b \frac{{\zeta (3)}}{{\pi ^2 }}T^3 {\rm{  }}\text{ and }{\rm{  }}m_{pion}  = \frac{1}{{{\rm{a}}_0 }}
\end{align*}
And for fermions(let them $p^+, p^-, n$) we approximate(ignoring the chemical potential)
\[
\varepsilon _f  = g_f \frac{7}{8}\frac{{\pi ^2 }}{{30}}T^4  \to \varepsilon _f  = g_f \frac{7}{8}\frac{{\pi ^2 }}{{30}}T^4  + m_f n_f {\rm{  }}\text{ with }{\rm{  }}n_f  = \frac{{N_f }}{V} = g_f \frac{{3\zeta (3)}}{{4\pi ^2 }}T^3
\]
So (3.15) becomes
\begin{align}
\frac{{\sigma _{{\rm{a}}^3 } {\rm{a}}}}{{4\pi {\rm{a}}^3 /3}} &= g_q \frac{{\mu _0^4 }}{{4 \cdot 2\pi ^2 }} = \varepsilon _f  + \varepsilon _b  + \varepsilon _{ph}  \nonumber\\
& = g_f \frac{7}{8}\frac{{\pi ^2 }}{{30}}T_c^4  + m_f n_f  + g_b \frac{{\pi ^2 }}{{30}}T_c^4  + \frac{1}{{{\rm{a}}_0 }}n_b  + g_{ph} \frac{{\pi ^2 }}{{30}}T_c^4
\end{align}
With $g_{quarks}=2_{flavor}\ast2_{charge}\ast2_{spin}\ast3_{color}$,  $g_f=3_{charge}\ast2_{spin}$ , $g_b=3_{charge}$ and $g_{ph}=2_{polarization}$ \\

Now we calculate (3.17) for $1/{\rm{a}_0}=135Mev(\pi^0)$, $\mu_0=285.15Mev$, and $T_c=124.9Mev$
we have
\begin{align*}
 2.0096*10^9 &Mev^4  = 6*\frac{7}{8}\frac{{\pi ^2 }}{{30}}({\rm{124}}{\rm{.9}})^4  + m_f 6*\frac{{3\zeta (3)}}{{4\pi ^2 }}({\rm{124}}{\rm{.9}})^3  \\
&+ 3*\frac{{\pi ^2 }}{{30}}({\rm{124}}{\rm{.9}})^4
  + 135*3*\frac{{\zeta (3)}}{{\pi ^2 }}({\rm{124}}{\rm{.9}})^3  + 2*\frac{{\pi ^2 }}{{30}}({\rm{124}}{\rm{.9}})^4
 \end{align*}
\\
Its solution is $m_f=1023Mev$ \\
We keep $2.0096*10^9Mev^4$ as smallest possible energy density.\\

For $1/{\rm{a}_0}=140Mev$ , $\mu_0= 295.7Mev$ so $T_c= 129.5Mev$ the mass $m_f$ becomes $m_f= 798.4Mev$. Therefore it must be $135{Mev}<1/{\rm{a}_0}<140{Mev}$. \\

For $1/{\rm{a}_0}= 136.8{Mev}$ we have $T_c=126.56Mev$ then the mass $m_f$ becomes $m_f\approx 938Mev$ so the fermions(hadrons) are the baryons$(p^+, p^-, n)$.\\

Therefore we fix it $1/{\rm{a}_0}= 136.8Mev$, we use it to cancel the dark matter. Maybe there is an external pressure $-P_{ex}$ so the lost energy is $P_{ex}4\pi{\rm{a}}^3/3$ . \\

Now we try to calculate the ratio $N_q/N_h$ . From the condensation relation
\[
N_q \delta \mu _q  + N_h \delta \mu _h  = 0
\]
$N_h$ is the hadrons(consider only the fermions) and $\mu_h$ is their chemical potential. \\

We assumed before the relation for the quarks chemical potential
\[
{\rm{ }}\mu (r) = \mu _0  + u(r){\rm{  }}\text{ with }{\rm{  }}u(r) =  - \frac{{\alpha _s }}{r}\left( {1 - e^{ - r/a} } \right)
\]
\[
{\rm{ }}\text{ so }{\rm{ }}\delta \mu _q (r) = u(r) =  - \frac{{\alpha _s }}{r}\left( {1 - e^{ - r/a} } \right)
\]
The effect of this changing  appeared in $y=\alpha_s/{\rm{ a}}\mu_0$  in the results.\\
For the hadrons we have
\[
\delta \mu _h  =  - \frac{{N_q }}{{N_h }}\delta \mu _q  =  - \frac{{N_q }}{{N_h }}u(r)
\]
That is right if we consider the hadrons are massless, that is when $T>>T_c$  and $y<<1$(in the condensation phase, below the curve $X_cY_c$) so we have the chemical potential for the hadrons
\[
{\rm{ }}\mu _h (r) = \mu _{0h}  - u(r){\rm{  }}\text{ with }{\rm{  }}u(r) =  - \frac{{\alpha _s }}{r}\left( {1 - e^{ - r/a} } \right)
\]
therefore we replace $y\rightarrow(-N_q\mu_{0q} / N_h\mu_{0h})y$ in the quarks energy to get the hadrons energy. \\
The energy of the hadrons becomes
\begin{eqnarray*}
E_{H,low}
      & =& \frac{1}{{1.9}}\frac{{7\pi ^2 }}{{240}}\mu _{0h}^4 g_h Vx^4 \left[ {3.78 + 2*x^{ - 2} \left( {0.82 + 1.16\frac{{N_q \mu _{0q} }}{{N_h \mu _{0h} }}y} \right.} \right.
\\
\\
   && \left. { + 0.41\left( {\frac{{N_q \mu _{0q} }}{{N_h \mu _{0h} }}} \right)^2 y^2 } \right) + x^{ - 4} \left( {0.08 + 0.23\frac{{N_q \mu _{0q} }}{{N_h \mu _{0h} }}y + 0.25\left( {\frac{{N_q \mu _{0q} }}{{N_h \mu _{0h} }}} \right)^2 y^2 } \right.
\\
\\
&&\text{ \text{ \text{ \text{ \text{ \text{ \text{ \text{  } } } } } } } }\left. {\left. { + 0.12\left( {\frac{{N_q \mu _{0q} }}{{N_h \mu _{0h} }}} \right)^3 y^3  + 0.02\left( {\frac{{N_q \mu _{0q} }}{{N_h \mu _{0h} }}} \right)^4 y^4 } \right)} \right]
\label{eq:multilineeq}
\end{eqnarray*}

assume $\mu _{0h}=\mu _{0q}$ and $y=1$ so
\begin{eqnarray*}
   E_{H,low}&=& \frac{1}{{1.9}}\frac{{7\pi ^2 }}{{240}}\mu _{0q}^4 g_h V \left[ {3.78x^4  + 2*x^2 \left( {0.82 + 1.16\frac{{N_q }}{{N_h }}} \right.} \right.\left. { + 0.41\left( {\frac{{N_q }}{{N_h }}} \right)^2 } \right)\\
\\
   && \text{ \text{ \text{ \text{  } } } }  \left. { + \left( {0.08 + 0.23\frac{{N_q }}{{N_h }} + 0.25\left( {\frac{{N_q }}{{N_h }}} \right)^2  + 0.12\left( {\frac{{N_q }}{{N_h }}} \right)^3  + 0.02\left( {\frac{{N_q }}{{N_h }}} \right)^4 } \right)} \right]{\rm{ }}
\label{eq:multilineeq}
\end{eqnarray*}
So the chemical potential $\mu_h$ of the hadrons becomes
\[
\mu _h^4  = \mu _{0q}^4 \left( {1 + \frac{{0.23}}{{0.08}}\left( {\frac{{N_q }}{{N_h }}} \right) + \frac{{0.25}}{{0.08}}\left( {\frac{{N_q }}{{N_h }}} \right)^2  + \frac{{0.12}}{{0.08}}\left( {\frac{{N_q }}{{N_h }}} \right)^3  + \frac{{0.02}}{{0.08}}\left( {\frac{{N_q }}{{N_h }}} \right)^4 } \right){\rm{ }}
\]
When $T<T_c$ the hadrons become massive, as we assumed before, so for massive hadrons with $m_f=938Mev$ we expect $\mu_h=m_f=938Mev$ when they cooled with small densities. Therefore
\[
\left( {{\rm{938}}} \right)^4  = \left( {{\rm{285}}{\rm{.15}}} \right)^4 \left( {1 + \frac{{0.23}}{{0.08}}\left( {\frac{{N_q }}{{N_h }}} \right) + \frac{{0.25}}{{0.08}}\left( {\frac{{N_q }}{{N_h }}} \right)^2  + \frac{{0.12}}{{0.08}}\left( {\frac{{N_q }}{{N_h }}} \right)^3  + \frac{{0.02}}{{0.08}}\left( {\frac{{N_q }}{{N_h }}} \right)^4 } \right){\rm{ }}
\]
Its positive solution is $N_q/N_h=3.1$ so they are the baryons( fermions with three quarks).\\
For 0 temperature fermions the chemical potential is approximated to
\[
\mu _0^2  = m^2  + \left( {\frac{N}{V}\frac{{6\pi ^2 }}{{g_f }}} \right)^{2/3}
\]
For low hadrons density we ignored the term
\[
\left( {\frac{N}{V}\frac{{6\pi ^2 }}{{g_f }}} \right)^{2/3}
\]

\section{The nuclear compression}
The cooled hadrons have high density, so there is hidden high pressure, that pressure makes influence $\delta\rm{a}$ so $\delta{y}$ near $y=1$ or it makes $y=1+\delta{y}$: $\delta{y}\approx0.005$ so the cooled quarks inside the hadrons fluctuate, this depends on the energy, if the energy is high then there are new hadrons.\\
This processes lets the interacted hadrons lose an kinetic energy and form the pions.

Because the number of quarks increases although the hadrons are fixed, therefore the hadrons energy decreases and they cannot spread away. We can see how the chemical potential of the interacted hadrons changes under the fluctuation $\delta{y}\sim\delta\rm{a}$( due to the quarks interaction) from the condensation relation   $N_q\delta\mu_q+N_h\delta\mu_h=0$  we have $\delta\mu_h={-N_q\delta\mu_q}/N_h$ \\
for the fluctuation $\delta{y}$ we have
\[
\delta \mu _h  =  - \frac{{N_q }}{{N_h }}\frac{{\partial \mu _q }}{{\partial y}}\delta y
\]
from quarks chemical potential(4.4), we find
\[
\frac{{\partial \mu _q }}{{\partial y}} \prec 0{\rm{ }}\text{ so }{\rm{ }} - \frac{{\partial \mu _q }}{{\partial y}} \succ 0
\]
therefore we have
\[
\delta \mu _h  = \frac{{N_q }}{{N_h }}\left( { - \frac{{\partial \mu _q }}{{\partial y}}} \right)\delta y \prec 0{\rm{ }}\text{ when }{\rm{ }}\delta y \prec 0
\]
which is the quarks compressing, when the hadrons collide together this lets to $\delta{y}<0$(compression) so the hadrons loss energy and new hadrons are created.\\
And when they try to extend(spread away) $\delta{y}>0$ so $\delta{\mu_h}>0$ , there will be a negative potential.\\

For the interacted hadrons pressure we have the phase changing relation $V_q \delta P_q  + V_h \delta P_h  = 0$ : V  volume, we have
\begin{equation*}
\delta P_h  =  - \frac{{V_q }}{{V_h }}\delta P_q  =  - \frac{{V_q }}{{V_h }}\frac{{\partial P_q }}{{\partial y}}\delta y
\end{equation*}
because ${\partial{P_q}}/{\partial{y}}<0 \rightarrow{-\partial{P_q}}/{\partial{y}}>0$  therefore when the hadrons collide together $\delta{y}<0$ so their pressure decreases, they lose energy, so new hadrons are created.\\

We have
\[
\delta y = \left( { - \frac{{V_q }}{{V_h }}\frac{{\partial P_q }}{{\partial y}}} \right)^{ - 1} \delta P_h {\rm{  }}\text{ at }{\rm{ }}y = 1
\]
So the hadrons chemical potential becomes
\[
\delta \mu _h  = \frac{{N_q }}{{N_h }}\left( { - \frac{{\partial \mu _q }}{{\partial y}}} \right)\left( { - \frac{{V_q }}{{V_h }}\frac{{\partial P_q }}{{\partial y}}} \right)^{ - 1} \delta P_h {\rm{  }}\text{ : }{\rm{ }}y = 1
\]
It becomes
\begin{equation}
\delta \mu _h  = \frac{{N_q V_h }}{{N_h V_q }}\left( {\frac{{\partial \mu _q }}{{\partial y}}} \right)\left( {\frac{{\partial P_q }}{{\partial y}}} \right)^{ - 1} \delta P_h {\rm{  }}\text{ : }{\rm{ }}y = 1
\end{equation}
We can relate this changing to a constant nuclear potential. Like to write
\begin{equation}
\delta \mu _h  =  - V_0
\end{equation}
$V_0$ is the potential for each hadron.\\

So when the hadron(fermions, like protons or neutrons) join, their density increases $\delta\mu_h>0$ so their pressure rises $\delta{P_h}>0$, therefore  there is a negative potential $V_0<0$ . At low energies this potential prevents them from spreading away.\\

Now we calculate
\[
\delta \mu _h  = \frac{{N_q V_h }}{{N_h V_q }}\left( {\frac{{\partial \mu _q }}{{\partial y}}} \right)\left( {\frac{{\partial P_q }}{{\partial y}}} \right)^{ - 1} \delta P_h {\rm{  }}\text{ : }{\rm{ }}y = 1
\]
We use the pressure at low energy(3.7)
\begin{equation*}
\begin{array}{l}
  P_{low}/{\mu _0^4 }= \left( {9 \cdot 1.9*240} \right)^{ - 1} 7\pi ^2 g_q \left[ {3 \cdot 3.78x^4  + 3*2*x^2 \left( {0.82 - 1.16y + 0.41y^2 } \right)} \right. \\
 \\
 \text{ \text{ \text{ \text{ \text{ \text{  } } } } } }\text{  }{\rm{ + 3*(}}0.08 - 0.23y + 0.25y^2  - 0.12y^3  + 0.02y^4 ){\rm{ + }}2yx^2 \left( { - 1.16 + 0.82y} \right) \\
 \\
 \text{ \text{ \text{ \text{ \text{ \text{ \text{ \text{ \text{ \text{ \text{  } } } } } } } } } } }\text{  }\left. { + y\left( { - 0.23 + 0.5y - 0.36y^2  + 0.08y^3 } \right)} \right] \\
 \end{array}
\end{equation*}
we get
\begin{equation}
\frac{{\partial P_q }}{{\partial y}} =  - \frac{{0.076*7*\pi ^2 *g_q \mu _0^4 }}{{240*3*1.9}}{\rm{  }}\text{ : }{\rm{   }}x_c  = 0.{\rm{438  }}\text{ and }{\rm{ }}y = 1
\end{equation}
Using the relation
\[
\mu _0  = \frac{1}{{{\rm{a}}_{\rm{0}} }}\left( {\frac{{8*9\pi }}{{g_q }}} \right)^{\frac{1}{3}}
\]
we have
\[
\begin{array}{l}
 \mu _0  = 285.15Mev\text{ for }g_q  = 24\text{ and }1/{\rm{a_0}}= 135Mev\text{ for }\pi ^0  \\
 \\
 \mu _0  = {\rm{295}}.{\rm{7}}Mev\text{ for }1/{\rm{a_0}} = 140Mev\text{ for }\pi ^ -  \text{ and }\pi ^ +   \\
 \end{array}
\]
So the chemical potential $\mu_0$ is in the range from $285.15Mev$  to $295.7Mev$ therefore
\begin{eqnarray*}
   \frac{{\partial P_q }}{{\partial y}}&=&    - 6.06*10^8 Mev^4 {\rm{  }}\text{ for }\mu_0={\rm{285.15}}Mev
\\
 \frac{{\partial P_q }}{{\partial y}} &=&  - 7.01*10^8 Mev^4 {\rm{  }}\text{ for }\mu_0={\rm{295.7}}Mev
\label{eq:multilineeq}
\end{eqnarray*}
Now we try to calculate $\partial \mu _q /\partial y$, according to low energy
\[
\begin{array}{l}
 E_{low}  = \left( {1.9*240} \right)^{ - 1} 7\pi ^2 \mu _0^4 g_q Vx^4 \left[ {3.78 + 2*x^{ - 2} \left( {0.82 - 1.16y + 0.41y^2 } \right)} \right. \\
 \\
 \text{ \text{ \text{  } } }\left. { + x^{ - 4} \left( {0.08 - 0.23y + 0.25y^2  - 0.12y^3  + 0.02y^4 } \right)} \right] \\
 \end{array}
\]
we can equivalent
\begin{equation}
\mu ^4  = \mu _0^4 \left( {1 - \frac{{0.23}}{{0.08}}y + \frac{{0.25}}{{0.08}}y^2  - \frac{{0.12}}{{0.08}}y^3  + \frac{{0.02}}{{0.08}}y^4 } \right)
\end{equation}
But $\partial\mu/\partial y\rightarrow\infty$ when $y\rightarrow1$ so we replace
\[
\frac{{\partial \mu _q }}{{\partial y}} \to \frac{{\mu _{y = 1}  - \mu _{y = 0} }}{{1 - 0}} = \frac{{0 - \mu _{y = 0} }}{{1 - 0}} =  - \mu _{y = 0}  =  - \mu _0
\]
Therefore we have
\begin{align*}
\delta \mu _h  = \frac{{N_q V_h }}{{N_h V_q }}\left( {\frac{{\partial \mu _q }}{{\partial y}}} \right)\left( {\frac{{\partial P_q }}{{\partial y}}} \right)^{ - 1} \delta P_h &=\frac{{N_q V_h }}{{N_h V_q }}\mu _0 \left( {{\rm{0}}{\rm{.09}}\mu _0^4 } \right)^{ - 1} \delta P_h \\
&=\frac{{N_q V_h }}{{N_h V_q }}\left( {{\rm{0}}{\rm{.09}}\mu _0^3 } \right)^{ - 1} \delta P_h {\rm{   }}
\end{align*}
So we have
\begin{align}
&\delta \mu _h  = \frac{{N_q V_h }}{{N_h V_q }}\left( {{\rm{0}}{\rm{.09}}\mu _0^3 } \right)^{ - 1} \delta P_h  = 4.7*10^{ - 7} \frac{{N_q V_h }}{{N_h V_q }}\delta P_h {\rm{  }}\text{ for }{\rm{ }}\mu _{\rm{0}} {\rm{ = 285}}{\rm{.15}}Mev\nonumber\\
&and \nonumber \\
&\delta \mu _h  = 4.2*10^{ - 7} \frac{{N_q V_h }}{{N_h V_q }}\delta P_h {\rm{  }}\text{ for }{\rm{ }}\mu _{\rm{0}} {\rm{ = 295}}{\rm{.7}}Mev
\end{align}
We use them to cancel the dark matter and dark energy.

\section{The Big Bang}
We assume there were two universal phases, high energies massless particles phase(let them the quarks plasma) and then the massive low energies particles(let them the Hadrons).

The first phase associated with high energy density(drops from infinity to finite), the time of that stage is $\tau: 0\rightarrow \rm{a}_0=1/(135-140)Mev$ then the massive hadrons phase began(the time $t: 0\rightarrow\infty$).

In both stages the highest universal expanding must not excess the light speed, for the first phase, high energies massless quarks phase, the density of the energy is the same in all space points so the universal expanding is the same in every point in the space, we let the speed of that expanding equals the light speed, therefore the Hubble parameter $H(t<{\rm{a}}_0)$ of this stage $t<{\rm{a}}_0$ is given by (5.2).\\

To find the Hubble parameter for the massive Hadrons phase $H(t>{\rm{a}}_0)$, we suggest the geometry transformation(5.3) in which the time $\tau: 0\rightarrow {\rm{a}}_0$ for the quarks corresponds to the time $t: 0\rightarrow\infty$ for the massive hadrons phase. We can relate that changing in the Geometry to the high differences in the energies densities of the two phases.\\
The phase $\tau: 0\rightarrow {\rm{a}}_0$ high quarks energy, uniform high energy density, massless,\ldots \\
The phase $t: 0\rightarrow\infty$  the massive hadrons, low energy density, separated particles,\ldots \\

Now we try to explain how the universe exploded and expanded, we start from our assumptions we made before and find the Hubble parameter and try to find the dark energy and matter.\\
We found that the quarks expand to the length $\rm{a}_0=1/(135-140)Mev$  then the hadrons appear instead. \\

We assume that the universe was created in every point in two dimensions space XY then the explosion in Z direction. That is by the quarks, in each point in XY flat the quarks were created and then they expanded in each point XY to the length ${\rm{a}}_0$ then the explosion in Z direction, the result is the universe in the space XYZ. \\
There was no universal explosion in the XY flat, the universal explosion was only in Z direction, in the flat XY there was extend due to the quarks expanding from $r =0$ to $r = {\rm{a}}_0=1/(135-140)Mev$ the flat XY was infinity before the quarks expanding and it is infinity after that expanding, what happened is increasing in the number of the XY points, then the explosion in Z direction.
We assume both expanding( XY and Z) occurred with the light speed c . \\

To find the lost matter, dark matter and dark energy, we use the relation(4.5) we found before:
\[
 \delta \mu _h  = \frac{{N_q V_h }}{{N_h V_q }}\left( {{\rm{0}}{\rm{.09}}\mu _0^3 } \right)^{ - 1} \delta P_h  = 4.7*10^{ - 7} \frac{{N_q V_h }}{{N_h V_q }}\delta P_h {\rm{  }}\text{ for }{\rm{ }}\mu _{\rm{0}} {\rm{ = 285}}{\rm{.15}}Mev
 \]
 and
\[
\delta \mu _h  = 4.2*10^{ - 7} \frac{{N_q V_h }}{{N_h V_q }}\delta P_h {\rm{  }}\text{ for }{\rm{ }}\mu _{\rm{0}} {\rm{ = 295}}{\rm{.7}}Mev
\]
Here we relate this changing in the pressure $\delta P$(independent on time) to the hadrons condensation process to form the nucleuses, where the global pressure $\delta P=\delta P_h$ extremely dropped due to the nuclear attractive potential (make it the nuclear binding energy) $V_0= (-7-8)Mev[3]$. This pressure $\delta P_h$ is remained contained in the nucleuses, but globally is not visible. \\

So there is hidden global pressure $\delta P_h$ and we have to include that problem in the Friedman equations solutions, we notice that the nuclear attractive potential lets to increasing in the cooled hadrons densities. Therefore the decreasing in the hadrons pressure associated with the increasing of their densities (inside the nucleuses).\\
The result is excess in the local energy density, that effects appear in the equations, that is, the matter density appears to be larger than the right energy density. So there is neither dark matter nor dark energy, it is just global and local densities.

We start from the defining the scale parameter $R(t)$ for the universe expanding, we write[6]
\begin{equation}
ds^2  =  - dt^2  + R^2 (t)\left( {\frac{{dr^2 }}{{1 - kr^2 }} + r^2 d\Omega ^2 } \right)
\end{equation}
We set $k=0$ flat Universe.
Now we try to find the Hubble parameter
\[
H(t) = \frac{1}{{R(t)}}\frac{{dR(t)}}{{dt}} = \frac{{\dot R(t)}}{{R(t)}}
\]
There are two phases $t<{\rm{a}_0}$ free quarks phase and $t>{\rm{a}_0}$ hadrons phase which is the expanding in Z direction.\\
That means there are two different spacetime Geometric, $t<{\rm{a}_0}$ and $t>{\rm{a}_0}$.\\

In the first phase $\tau=t<{\rm{a}_0}$ the expanding is the same in all space points, so the expanding velocity
\[
\frac{{dR_1 }}{{dt}} = \dot R(t)r
\]
is the same in all space points and equals the light speed $c=\hbar=1$ here, so
\[
1 = \dot R(t)r\text{ :\text{  } }t < {\rm{a}} = {\rm{a}}_0
\]
Therefore
\[
\dot R(t) = \frac{1}{r}\text{ :\text{  } }t < {\rm{a}} = {\rm{a}}_0
\]
So we can write
\[
R(t) = \frac{t}{r}\text{ :\text{  } }t < {\rm{a}} = {\rm{a}}_0
\]
So the Hubble parameter becomes
\begin{equation}
H(t) = \frac{{\dot R(t)}}{{R(t)}} = \frac{{1/r}}{{t/r}} = \frac{1}{t}\text{ :\text{  } }t < {\rm{a}} = {\rm{a}}_0
\end{equation}
Now we want to find the Hubble parameter in the phase $t>{\rm{a}_0}$  low energy phase.\\
Actually when the quarks expand from $r =0$ to $r ={\rm{a}}\rightarrow{\rm{a}_0}$ there will be infinity points expanding, so infinity expanding distance in XY space, but the expanding cannot excess the light speed $c=1$ therefore an explosion occurred in Z direction, so the universal explosion.\\
Therefore the time $t =\tau \text{ : }0\rightarrow \rm{a}_0$ for  the free quarks phase will associate with $t\text{ : } 0\rightarrow\infty$ for the universal expanding, so we make the geometry transformation
\begin{equation}
t = \frac{{ - c_0 }}{{\tau  - {\rm{a}}_0 }}\text{ :\text{  } }\tau  < {\rm{a}}_0
\end{equation}
 $c_0$ is constant, we can relate that relation to a difference in spacetime Geometry. That means if the quarks space $r<{\rm{a}}_0=1/(135-140)Mev$ is flat, so the hadrons space is not, it is curved space, where we live.\\
It is convenient to consider the quarks space( $r<{\rm{a}}_0$ large energy density) is curved not our space(low energy density).\\

Now we can find the Hubble parameter for the universe $t\text{ : } 0\rightarrow\infty$ \\
we can find the Hubble parameter $H(t>{\rm{a}}_0)$ for the geometry $t\text{ : } 0\rightarrow\infty$ from $H(t<{\rm{a}}_0)$ :
\[
H(\tau  < {\rm{a}}) = \frac{1}{{R(\tau  < {\rm{a}})}}\frac{{dR(\tau  < {\rm{a}})}}{{d\tau }} = \frac{1}{{f(r,\theta ,\varphi )R(t > {\rm{a}})}}\frac{d}{{d\tau }}f(r,\theta ,\varphi )R(t > {\rm{a}}){\rm{ }}
\]
We set the geometry transformation
\[
R(\tau<{\rm{a}})=f(r,\theta, \varphi)R(t>{\rm{a}})\]
so
\begin{equation}
\frac{1}{\tau } = \frac{1}{{f(r,\theta ,\varphi )R(t > {\rm{a}})}}\frac{d}{{d\tau }}f(r,\theta ,\varphi )R(t > {\rm{a}}) = \frac{1}{{R(t > {\rm{a}})}}\frac{d}{{d\tau }}R(t > {\rm{a}})
\end{equation}
or
\[
\frac{1}{\tau } = \frac{1}{{R(t > {\rm{a}})}}\frac{{dt}}{{d\tau }}\frac{d}{{dt}}R(t > {\rm{a}}) = \frac{{dt}}{{d\tau }}\frac{1}{{R(t > {\rm{a}})}}\frac{d}{{dt}}R(t > {\rm{a}}) = \frac{{dt}}{{d\tau }}H(t > {\rm{a}})
\]
Using the geometry transformation
\[
t = \frac{{ - c_0 }}{{\tau  - {\rm{a}}_0 }}\text{ :\text{  } }\tau  < {\rm{a}} \to {\rm{a}}_0
\]
We have the Hubble parameter of the low energy density of cold Universe
\[
H(t > {\rm{a}}_0 ) = \frac{1}{{R(t > {\rm{a}}_0 )}}\frac{d}{{dt}}R(t > {\rm{a}}_0 ) = \frac{{c_0 }}{{t{\rm{(a}}_0 t - c_0 )}} = \frac{1}{{t\left( {\frac{{{\rm{a}}_0 }}{{c_0 }}t - 1} \right)}}{\rm{ }} = \frac{1}{{t\left( {c'_0 t - 1} \right)}}
\]
${c'}_0$ is constant. \\

The  Friedman equations can be written, for $k=0$, like[6]
\[
3\frac{{\dot R^2 (t)}}{{R^2 (t)}} = 8\pi G_N \rho  + \Lambda \text{ \text{ \text{  } } }\text{ \text{ \text{ (1) } } }
\]
\begin{equation}
 - \frac{{\ddot R(t)}}{{R(t)}} + \frac{{\dot R^2 (t)}}{{R^2 (t)}} = 4\pi G_N (\rho  + p)\text{ \text{ \text{  } } }{ \text{ \text{ (2) } } }
\end{equation}
\[
\frac{d}{{dt}}(\rho  + \delta p) =  - 3(\rho  + p)\frac{{\dot R(t)}}{{R(t)}}\text{ \text{ \text{  } } }{ \text{ \text{ (3) } } }
\]
To control(or cancel) the dark matter and energy, we make the transformations in the Friedman equations which keep the Hubble parameter unchanged
\[
3\frac{{\dot R^2 (t)}}{{R^2 (t)}} = 8\pi G_N (\rho  + \delta P) + \Lambda  - 8\pi G_N \delta P \text{ \text{ \text{  } } }\text{ \text{ \text{ (1') } } }
\]
\begin{equation}
 - \frac{{\ddot R(t)}}{{R(t)}} + \frac{{\dot R^2 (t)}}{{R^2 (t)}} = 4\pi G_N (\rho  + \delta P + p - \delta P)\text{ \text{ \text{  } } }\text{ \text{ \text{ (2') } } }
\end{equation}
\[
\frac{d}{{dt}}(\rho  + \delta p) =  - 3(\rho  + \delta P + p - \delta P)\frac{{\dot R(t)}}{{R(t)}} \text{ \text{  } }\text{ \text{ \text{  } } }\text{ \text{ \text{ (3') } } }
\]
So we have(for same Hubble parameter we had before)
\[
\rho ' = \rho  + \delta p_h
\]
\[
p' = p - \delta p_h
\]
\[
\Lambda ' = \Lambda  - 8\pi G_N \delta p_h  = 0
\]
For the universal nuclear condensation, we assume the universal change $\delta\rho=\delta P=\delta p_h>0$ is independent on the time.\\

We can say $\rho', p' \text{ and } \Lambda' =0, P'=0$ are for the located matter, when the hadrons are cooled, they condense and locate in small volumes with high matter density, because of the strong nuclear attractive interaction, so their pressure extremely decreases $P'\approx0$. That pressure is contained(hidden) in the nucleus.\\
It is like to condense a gas with certain mass $m$ and fixed volume $V$, the density $m/V$ is the same before and after the condensation, but the real density of the produced liquid is not. Like that we consider $\rho$  the right matter $\rho_{matter}$ and the problems; the increasing $\rho ' = \rho  + \delta p_h$ and $ \Lambda \neq0$ are because of the phase changing.

We set $\rho'=\rho(t)$ and solve the two equations:
\[
 - \frac{{\ddot R(t)}}{{R(t)}} + \frac{{\dot R^2 (t)}}{{R^2 (t)}} = 4\pi G_N \rho (t)\text{ \text{ \text{  } } }\text{ \text{ \text{ (2') } } }
\]
\[
\frac{d}{{dt}}(\rho  + \delta p) = \dot \rho (t) =  - 3\rho (t)\frac{{\dot R(t)}}{{R(t)}}\text{ \text{ \text{  } } }\text{ \text{ \text{ (3') } } }
\]
using the Hubble parameter
\[
H(t) = \frac{1}{R}\frac{{dR}}{{dt}} = \frac{1}{{t\left( {c'_0 t - 1} \right)}}\text{ \text{ \text{  } } }\text{ \text{ \text{ : } } }t > {\rm{a}}_0
\]
From (3') we have
\[
\frac{{ - 1}}{3}\frac{{R(t)}}{{\dot R(t)}}\dot \rho (t) = \rho (t)
\]
so (2') becomes
\[
 - \frac{{\ddot R(t)}}{{R(t)}} + \frac{{\dot R^2 (t)}}{{R^2 (t)}} = \frac{{ - 4\pi G_N }}{3}\frac{{R(t)}}{{\dot R(t)}}\dot \rho (t)
\]
This equation becomes
\[
\frac{{\dot R(t)}}{{R(t)}}\left( { - \frac{{\ddot R(t)}}{{R(t)}} + \frac{{\dot R^2 (t)}}{{R^2 (t)}}} \right) = \frac{{ - 4\pi G_N }}{3}\dot \rho (t){\rm{  }}
\]
or
\[
H(t)\left( { - \frac{{\ddot R(t)}}{{R(t)}} + H^2 (t)} \right) = \frac{{ - 4\pi G_N }}{3}\dot \rho (t)
\]
Using
\[
\frac{d}{{dt}}\frac{{\dot R(t)}}{{R(t)}} = \frac{{\ddot R(t)}}{{R(t)}} - \frac{{\dot R^2 (t)}}{{R^2 (t)}}
\]
we get
\[
H(t)\frac{d}{{dt}}H(t) = \frac{{4\pi G_N }}{3}\dot \rho (t){\rm{  }} \to \frac{1}{2}H(t)^2  = \frac{{4\pi G_N }}{3}\left( {\rho (t) - \rho _0 } \right)
\]
For finite results we put $\rho_0=0$ so
\[
\frac{1}{2}H(t)^2  = \frac{{4\pi G_N }}{3}\rho (t)
\]
Now we calculate the contributions of the vacuum energy to the total energy using the cosmological constant $\Lambda'$ from (1')
\[
\Omega _{\Lambda'}   = \frac{{\rho _\Lambda'  }}{{\rho _c }} = \frac{\Lambda' }{{3H^2 }} = \frac{{3H^2  - 8\pi G_N \rho (t)}}{{3H^2 }} = 1 - 2\frac{{4\pi G_N }}{{3H^2 }}\rho (t) = 1 - 2\frac{1}{{H^2 }}\frac{1}{2}H(t)^2  = 0
\]
with the critical energy density
\[
\rho _c  = \frac{{3H^2 }}{{8\pi G_N }}
\]
So the vacuum energy density is canceled, and the total energy is the matter energy $\Omega_{matter}=1$ so $\rho(t)/\rho_c =1$ .

Here $\rho(t)=\rho_c$ is $\rho(t)=\rho'=\rho_{matter}+\delta p_h$, So $\rho(t)$ is higher than the right matter $\rho_{matter}$ . \\

Now we see if this relation is satisfied or not.
We use the global change on the pressure $\delta p=\delta p_h>0$ which we derived(4.5):
\[
\delta \mu _h  = \frac{{N_q V_h }}{{N_h V_q }}\left( {{\rm{0}}{\rm{.09}}\mu _0^3 } \right)^{ - 1} \delta P_h  = 4.7*10^{ - 7} \frac{{N_q V_h }}{{N_h V_q }}\delta P_h {\rm{  }}\text{ \text{ \text{ for } } }{\rm{ }}\mu _{\rm{0}} {\rm{ = 285}}{\rm{.15}}Mev
\]
and
\[
\delta \mu _h  = 4.2*10^{ - 7} \frac{{N_q V_h }}{{N_h V_q }}\delta P_h {\rm{  }}\text{ \text{ \text{ for } } }{\rm{ }}\mu _{\rm{0}} {\rm{ = 295}}{\rm{.7}}Mev{\rm{ }}
\]
Now we try to find $V_q /V_h$  the quarks volume $V_q=Sd_q$ and the hadrons volume $V_h=Sd_h$ as the figure
\begin{figure}[h!]
  \includegraphics[width=0.55\textwidth]{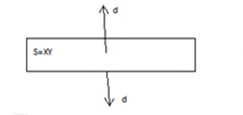}
  \caption{The universal explosion in Z direction starting from XY flat.}
\end{figure}

Where the universal explosion in the $z=d$ direction.\\
If we assume the explosion speed is the same for both hadrons and quarks, light speed $\nu=c=1$, so for the quarks
\[
H_q (t) = \frac{{\dot R(t)_q }}{{R(t)_q }} = \frac{{\nu _q }}{{d_q }} = \frac{1}{{d_q }}
\]
For the hadrons
\[
H_h (t) = \frac{{\dot R(t)_h }}{{R(t)_h }} = \frac{{\nu _h }}{{d_h }} = \frac{1}{{d_h }}
\]
therefore
\[
\frac{{V_q }}{{V_h }} = \frac{{Sd_q }}{{Sd_h }} = \frac{{d_q }}{{d_h }} = \frac{{H_h }}{{H_q }}
\]
$H_h$ is the universal Hubble parameter, today is
\begin{align*}
H = 71km/s/mpc = 2.3 \cdot 10^{ - 18} s^{ - 1}  &= 2.3 \cdot 10^{ - 18}  \cdot 6.58 \cdot 10^{ - 22} Mev\\
& = 151.34 \cdot 10^{ - 41} Mev
\end{align*}
The quarks Hubble parameter $H_q=1/\tau {\rightarrow}1/{\rm{a}_0}=(135-140)Mev$  \\
So we have( for 135Mev)
\[
\frac{{H_h }}{{H_q }} = \frac{{151.34 \cdot 10^{ - 41} Mev}}{{135Mev}} = {\rm{1}}{\rm{.127}} \cdot 10^{ - 41}
\]
Therefore
\[
\frac{{V_q }}{{V_h }} = \frac{{H_h }}{{H_q }} = {\rm{1}}{\rm{.127}} \cdot 10^{ - 41}
\]
We set $\delta\mu_h=-V_0 =(7-8)Mev$  the nuclear potential(nucleon binding energy). \\
Therefore, from (4.5), we have
\begin{align*}
&\delta \rho  = \delta P_h  =  - \frac{{N_h }}{{N_q }} \cdot {\rm{1}}{\rm{.127}} \cdot 10^{ - 41}  \cdot \frac{{ - V_0 }}{{47}} \cdot 10^8 Mev^4 {\rm{ }}\\
&for\text{  }{\rm{ }}\frac{1}{{{\rm{a}}_0 }} = 135Mev:{\rm{ }}\mu _{\rm{0}} {\rm{ = 285}}{\rm{.15}}Mev
\end{align*}
and
\begin{align*}
&\delta \rho  = \delta P_h  =  - \frac{{N_h }}{{N_q }} \cdot {\rm{1}}{\rm{.087}} \cdot 10^{ - 41}  \cdot \frac{{ - V_0 }}{{42}} \cdot 10^8 Mev^4 {\rm{ }}\\
&for\text{  }{\rm{ }}\frac{1}{{{\rm{a}}_0 }} = 140Mev:{\rm{ }}\mu _{\rm{0}} {\rm{ = 295}}{\rm{.7}}Mev
\end{align*}
For $N_h/N_q=1/5$ , like the interaction $P^+ +\pi^-\rightarrow n$ the neutron n appears to have five quarks, that is acceptable according to the fields dual behavior.\\
Therefore
\begin{align*}
&\delta \rho  = \delta P_h  =  - \frac{1}{5} \cdot {\rm{1}}{\rm{.127}} \cdot 10^{ - 41}  \cdot \frac{{ - 7}}{{47}} \cdot 10^8 Mev^4 {\rm{ = 335}}{\rm{.7*}}10^{ - 37} Mev^4 {\rm{ }}\\
&\text{ for }{\rm{ }}\mu _{\rm{0}} {\rm{ = 285}}{\rm{.15}}Mev{\rm{ }}\text{ and }{\rm{ }}V_0  =  - 7Mev
\end{align*}
and
\begin{align*}
&\delta \rho  = \delta P_h  =  - \frac{1}{5} \cdot {\rm{1}}{\rm{.087}} \cdot 10^{ - 41}  \cdot \frac{{ - 8}}{{42}} \cdot 10^8 Mev^4 {\rm{ = 414*}}10^{ - 37} Mev^4 {\rm{ }}\\
&\text{ for }{\rm{ }}\mu _{\rm{0}} {\rm{ = 295}}{\rm{.7}}Mev{\rm{ }}\text{ and }{\rm{ }}V_0  =  - 8Mev
\end{align*}
So the change $\delta\rho=\delta{P_h}$ is in the range:
 \[ \text{ from }{\rm{ 335}}{\rm{.7*}}10^{ - 37} Mev^4 {\rm{ }}\text{ to }{\rm{ 414*}}10^{ - 37} Mev^4\]
Therefore the visible matter is in the range
\[
\text{ from }\rho _{matter} {\rm{ = }}\rho _c  - \delta p_h  = {\rm{ 335}}{\rm{.7*}}10^{ - 37} Mev^4 {\rm{ }}
\]
\[
\text{ to }\rho _{matter} {\rm{ = }}\rho _c  - \delta p_h  = {\rm{414*}}10^{ - 37} Mev^4
\]
\\
For the critical energy $\rho _c  = 406 \cdot 10^{ - 37} Mev^4$ the visible matter is in the range
\[
\text{ from }\rho _{matter}  = 0{\rm{ }}\text{ to }{\rm{ }}\rho _{matter}  = {\rm{70}} \cdot 10^{ - 37} Mev^4
\]
The right baryonic matter energy density is
\[
\rho _b  = 4.19 \cdot 10^{ - 31} g/cm^3  \approx 17.97 \cdot 10^{ - 37} Mev^4
\]
Which belongs to the range $0$ to $70*10^{-37}Mev^4$ \\
We can control this and have
 \[\rho _{matter} {\rm{ = }}\rho _c  - \delta P_h  = 406{\rm{*}}10^{ - 37}  - \delta P_h  = 17.97 \cdot 10^{ - 37} Mev^4\]
by finding r:
\[
140r + 135(1 - r) = \frac{1}{{{\rm{a}}_0 }}
\]
With $1/{\rm{ a }_0}$ satisfies
\[
406{\rm{*}}10^{ - 37} Mev^4  - \delta P_h  = 17.97 \cdot 10^{ - 37} Mev^4
\]
For $1/{\rm{ a }_0}=136.8Mev$( we used it in 3.17 to have $m_f\approx938Mev$), the chemical potential becomes $\mu_0=288.95Mev$. And with $V_0=7.776Mev$ we get
\[
\delta \rho  = \delta P_h {\rm{ = 335}}{\rm{.7*}}10^{ - 37} *\left( {\frac{{{\rm{288}}{\rm{.95}}}}{{{\rm{285}}{\rm{.15}}}}} \right)^3 *\frac{{{\rm{7}}{\rm{.776}}}}{7}Mev^4 {\rm{ = 388*}}10^{ - 37} Mev^4
\]
The matter density becomes
\[
\rho _{matter} {\rm{ = }}406 \cdot 10^{ - 37} Mev^4  - {\rm{ 388*}}10^{ - 37} Mev^4  = {\rm{17}}{\rm{.9}}Mev^4 {\rm{ }}
\]
Which is the right matter(global visible matter density).\\
Therefore we can control the  dark matter and dark energy. We can cancel them\\
Notice: Not all of those Ideas are contained in the references.

\end{document}